\documentclass[a4paper,twoside,fleqn]{aa}
\usepackage{graphics}
\usepackage{epsf}
\usepackage{natbib,epsfig,ifthen,graphics,lscape,flscape,amsmath,rotating, array, dcolumn}
\bibpunct{(}{)}{;}{a}{}{,}

\newcommand{\sun}{\hbox{$\odot$}}
\newcommand{\msun}{\hbox{M$_{\odot}$}}

\newcommand{\co}{\hbox{$^{12}\mathrm{CO}$}}
\newcommand{\kms}{\hbox{km\,s$^{-1}$}}
\newcommand{\wco}{\hbox{W$\mathrm{_{CO}}$}}

\newcommand{\ndash}{\hbox{\nobreakdash\textendash}}

\newcommand{\ssp}{\def\baselinestretch{1.0}\large\normalsize}

\newcommand{\hsp}{\def\baselinestretch{1.3}\large\normalsize}

\newcommand{\hts}{\vspace*{0.3ex}}

\begin{document}

\thesaurus{()}

\title{A Uniform CO Survey of the Molecular Clouds in Orion and Monoceros.}

\author{B.\,A.~Wilson
          \inst{1,2}
	  \and T.\,M.~Dame\inst{2}
	  \and M.\,R.\,W.~Masheder\inst{1}
	  \and P.~Thaddeus\inst{2}
          }

\offprints{T.\,M.~Dame tmdame@cfa.harvard.edu}

\institute{University of Bristol, Tyndall Avenue, Bristol, UK
         \and
             Harvard-Smithsonian Center for Astrophysics
             }

\date{Received December 23, 2003; accepted August 30, 2004}

\maketitle

\begin{abstract}

We report the results of a new large scale survey of the
Orion-Monoceros complex of molecular clouds made in the $J = 1
\rightarrow 0$ line of \co\ with the Harvard-Smithsonian 1.2\,m
millimetre-wave telescope.  The survey consists of 52\,288 uniformly
spaced spectra that cover an area of 432\,deg$^2$ on the sky and
represent the most sensitive large-scale survey of the region to date.
Distances to the constituent molecular clouds of the complex,
estimated from an analysis of foreground and background stars, have
provided information on the three dimensional structure of the entire
complex.


\end{abstract}

%

\section{Introduction}

Orion-Monoceros is one of the most important and best studied regions
of star-formation.  Much of its importance derives from its
favourable position on the sky and its proximity: it is located $\sim
15\degr$ below the Galactic plane and toward the outer Galaxy --- well
away from the potentially confusing molecular gas of the plane and the
Galactic centre; at a distance of $\sim 450$\,pc, it is relatively
close, and its constituent clouds can be observed with good linear
resolution even with a radio telescope of modest size.  The radial
velocities, proper motions and parallaxes of the stars associated with
the molecular gas are observable, and therefore the distances to the
clouds and the three-dimensional structure of the
complex can be estimated.

\begin{figure*}
\begin{center}
\leavevmode
\epsfxsize=18.0cm
\epsfbox{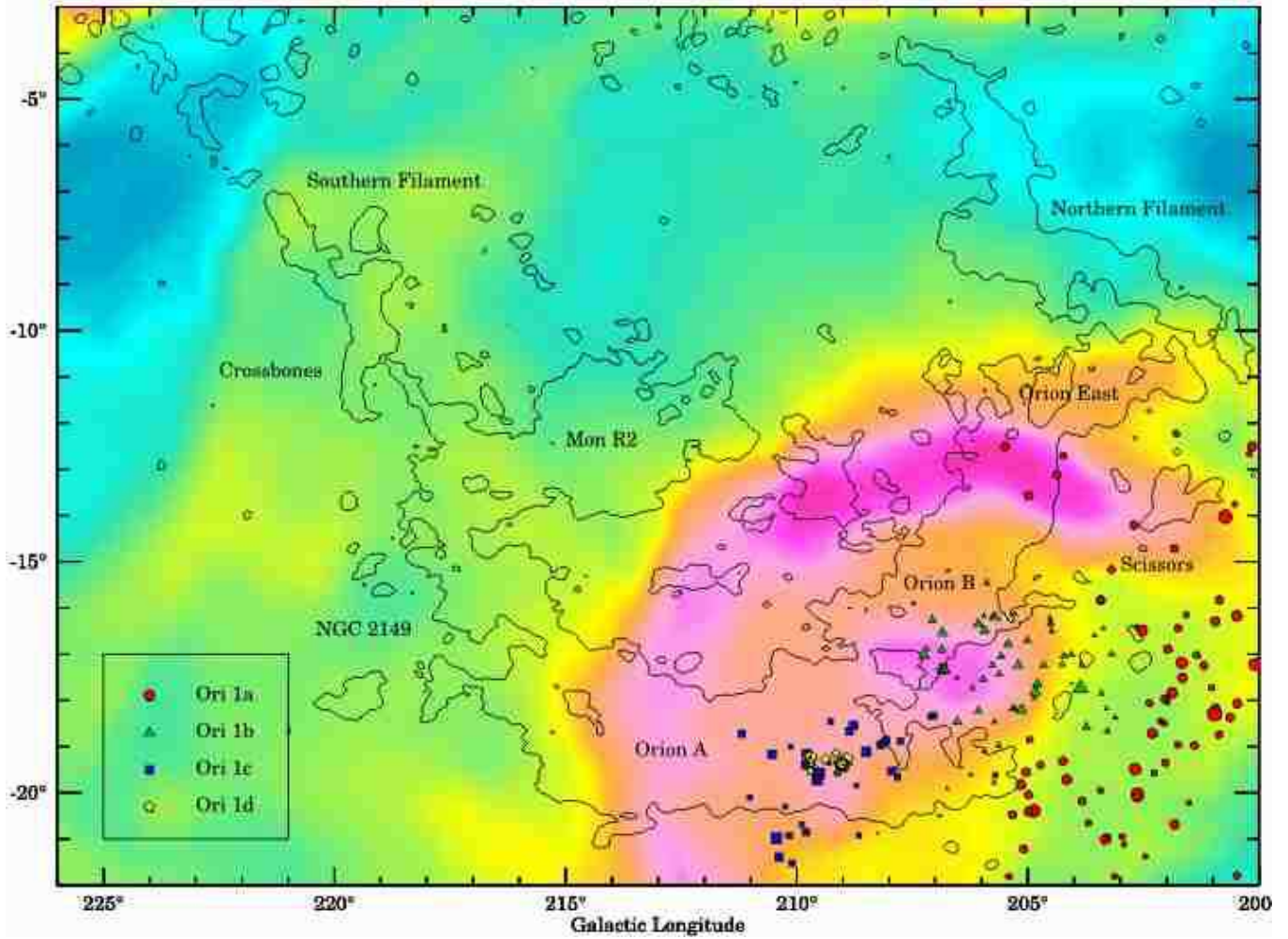}
\end{center}
\caption{Finder diagram for Orion-Monoceros. The colourscale represents the integrated intensity of ionized gas as traced by the WHAM H$\alpha$ survey (Reynolds et al. 1998). The bright crescent of emission is Barnard's loop which delineates the optical boundary of the Orion-Eridanus superbubble.  The positions of O and B stars from the Ori OB1 association (Brown et al. 1995) are plotted with symbols the sizes of which are proportional to the luminosity of the stars.  The four different symbols correspond to the four Ori OB\,1 subgroups. The boundary of the Orion-Monoceros molecular complex is marked by the 2.06\,K\,\kms\ (3$\sigma$) contour of the new CO survey.}\label{orim1}
\end{figure*}

Orion-Monoceros provides an excellent laboratory for studying the
interaction between massive stars and the interstellar medium (ISM).
Over the course of the last 12\,Myr the molecular clouds in the region
have been shaped, compressed and disrupted by the powerful ionizing
radiation, stellar winds and supernova (SN) explosions of the young
massive stars in the Orion OB association (Cowie, Songailia \& York
1979; Bally et al. 1987).\nocite{Cowie79}\nocite{Bally87} Compression of the clouds
may have triggered bursts of star-formation (Elmegreen \& Lada
1977)\nocite{Elmegreen77} that are traced by the four semi-distinct
Ori OB1 subgroups. The oldest subgroup, OB\,1a, lies in a region about
10\degr\ from the main clouds long-ago cleared of molecular gas, while
the youngest OB\,1d is currently forming the stars in the Orion
nebulae on the near side of the Orion~A cloud.  As well as disrupting
the molecular gas in Orion-Monoceros, the stars of Ori OB1 have also
inflated a huge elongated bubble of hot X-ray emitting gas, the
Orion-Eridanus supershell (Heiles, Haffner \& Reynolds
1999),\nocite{Heiles99} which is centered on the Orion star-forming
region 15\degr--20\degr\ ($\sim 150$\,pc) below the Galactic Plane.
As the bubble has expanded it has swept up a massive shell of dust and
gas that has become an optically bright crescent (Barnard's Loop) as
it moved into the steep density and pressure gradients associated with
the Galactic gas layer.  Fig.~\ref{orim1} shows a finder diagram for
Orion-Monoceros. The colourscale represents the integrated H$\alpha$
emission as measured by the Wisconsin H$\alpha$ mapper (WHAM --
Reynolds et al. 1998),\nocite{Reynolds98} and the contours show the
boundaries of the main molecular clouds as traced by the CO survey
presented here.  The positions of the O and B stars from the four
Ori-OB\,1 subgroups, from Brown et al. (1995),\nocite{Brown95} are
also plotted.

The first observations of molecular emission from Orion-Monoceros,
which were also the first detections of interstellar CO, were made toward
the Orion nebula by Wilson, Jefferts \& Penzias (1970)
\nocite{Wilson70} using the $J = 1 \rightarrow 0$ line of the normal
isotopic species. Subsequent studies using this line by Kutner et
al. (1977),\nocite{Kutner77} Chin (1977),\nocite{Chin77} and in
particular Maddalena et al. (1986)\nocite{Maddalena86} were conducted
primarily to determine the spatial extent of the complex.  It was
found that the complex contains three giant ($10^5$\msun) molecular
clouds (GMCs), Orion~A, Orion~B and Mon R2, two long
filaments which extend approximately 10\degr\ from the cloud complex
to the Galactic plane, and numerous smaller clouds.  Parts of
the complex have subsequently been observed in CO at higher resolution
either using the $J = 1 \rightarrow 0$ line or other isotopes or
transitions.  Most importantly, the majority of Mon~R2 was mapped in CO
($J = 1 \rightarrow 0$) by Xie \& Goldsmith (1994), \nocite{Xie94}the
central part of Orion~A was mapped using $^{13}$CO ($J = 1 \rightarrow
0$) by Bally et al. (1987)\nocite{Bally87}, by Nagahama et
al. (1998),\nocite{Nagahama98} and in the $J = 1 \rightarrow 0$ and 
$J = 2 \rightarrow 1$ lines of 
\co, $^{13}$CO and C$^{18}$O by Castets et al. (1990). 
\nocite{Castets90}In addition, the 
southern part of Orion B was mapped
in CO and $^{13}$CO ($J = 2 \rightarrow 1$ and $J = 3\rightarrow 2$)
by Kramer, Stutzki \& Winnewisser (1996), \nocite{Kramer96}and the
Orion~A and B clouds were mapped using the $J = 2 \rightarrow 1$ line
of CO (Sakamoto et al. 1994).  \nocite{Sakamoto94}However, to date
the only survey which has covered the whole region has been that of
Maddalena et al. (1986), a survey which is non-uniform in resolution,
sensitivity and sampling, and is heavily undersampled in many
places.

This paper presents a new CO ($J = 1 \rightarrow 0$) survey of the
Orion-Monoceros complex which consists of 52\,288 spectra uniformly
spaced in Galactic longitude and latitude. It is generally four times
more sensitive than the Maddalena et al. (1986) survey, since it
contains four times as many spectra, each with lower rms noise.  The
new observations cover a total area of more than 400\,sq\,deg on the
sky and map all of the major clouds of the Orion-Monoceros complex
with the exception of the $\lambda$-Orionis ring, which was the
subject of a separate study with the 1.2\,m telescope (Lang et
al. 2000).\nocite{Lang00}

This paper is organised as follows: In \S\ref{obs} the observations and data
reduction procedure are described. In \S\ref{sdist} the methods  used to estimate the masses
of and the distances to specific regions in the Orion-Monoceros complex are outlined.  
In \S\ref{sReg} these specific regions are described in detail. \S\ref{large} presents an overview of the large scale structure of the complex and describes a possible formation scenario. A summary and conclusions are presented in \S\ref{conc}.

\begin{figure*}
\begin{center}
\leavevmode
\epsfxsize=18cm
\epsfbox{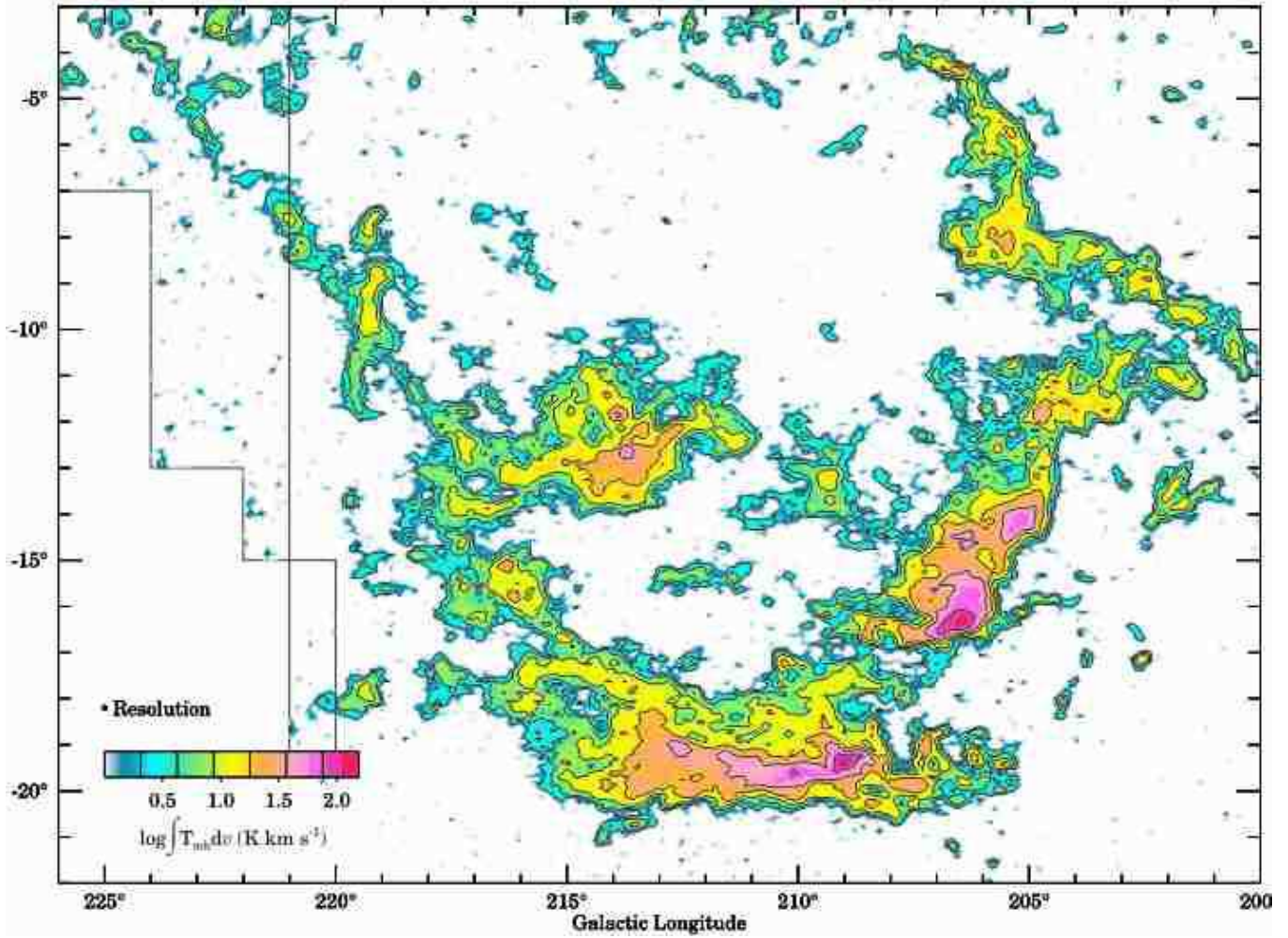}
\end{center}
\caption{Log velocity-integrated $(J = 1 \rightarrow 0)$ CO map of Orion-Monoceros. The integration range is $-3.0 < v_{\mathrm{LSR}} < 19.5$\,\kms.  Contours start at 0.313 (3$\sigma$) and go up in steps of 0.313\,Log(K\,\kms).  The dashed line marks the limit of the full-resolution survey.  The region on the high longitude side of the dashed line has been sampled with beamwidth spaced observations; that on the low longitude side has been observed twice with the second set of observations offset by half a beamwidth in longitude and latitude.} \label{orim2}
\end{figure*}

\section{Observations}\label{obs}

The 1.2\,m millimetre wave telescope at the Harvard-Smithsonian Center
for Astrophysics was used for all the $J = 1 \rightarrow 0$ \co\
observations presented here.  This telescope has a beamwidth (FWHM) of
8\farcm4 and a velocity resolution of 0.65 km\,$^{-1}$ at
115.2712\,GHz, provided by a 256 channel filter bank.  Spectral line
intensities were calibrated against an ambient temperature blackbody,
a standard Eccosorb chopper wheel (Davis et al 1973),\nocite{davis73}
which was rotated in front of the feed horn at 20\,Hz for 1\,s before
each observation.  All line intensities reported here are
in units of main beam brightness temperature $\mathrm{T_{mb} =
T^{*}_{A}/}\eta$, where $\mathrm{T^{*}_{A}}$ is the chopper calibrated
antenna temperature (Penzias et al. 1973)\nocite{penzias73} (i.e. the
antenna temperature corrected for atmospheric attenuation, ohmic
losses and rearward spillover), and $\eta$ is the main beam
efficiency, 0.82.  The emission in Orion is so strong and extensive that
Tmb will at some locations overestimate the intensity in the main beam
by a factor up to ~5\% owing to emission in the near sidelobes.
The detector was a liquid-helium cooled SIS
receiver with a single sideband noise temperature of $\sim 75$\,K (Pan
1983).\nocite{Pan83}

\begin{figure*}
\begin{center}
\leavevmode
\epsfxsize=8.75cm
\epsfbox{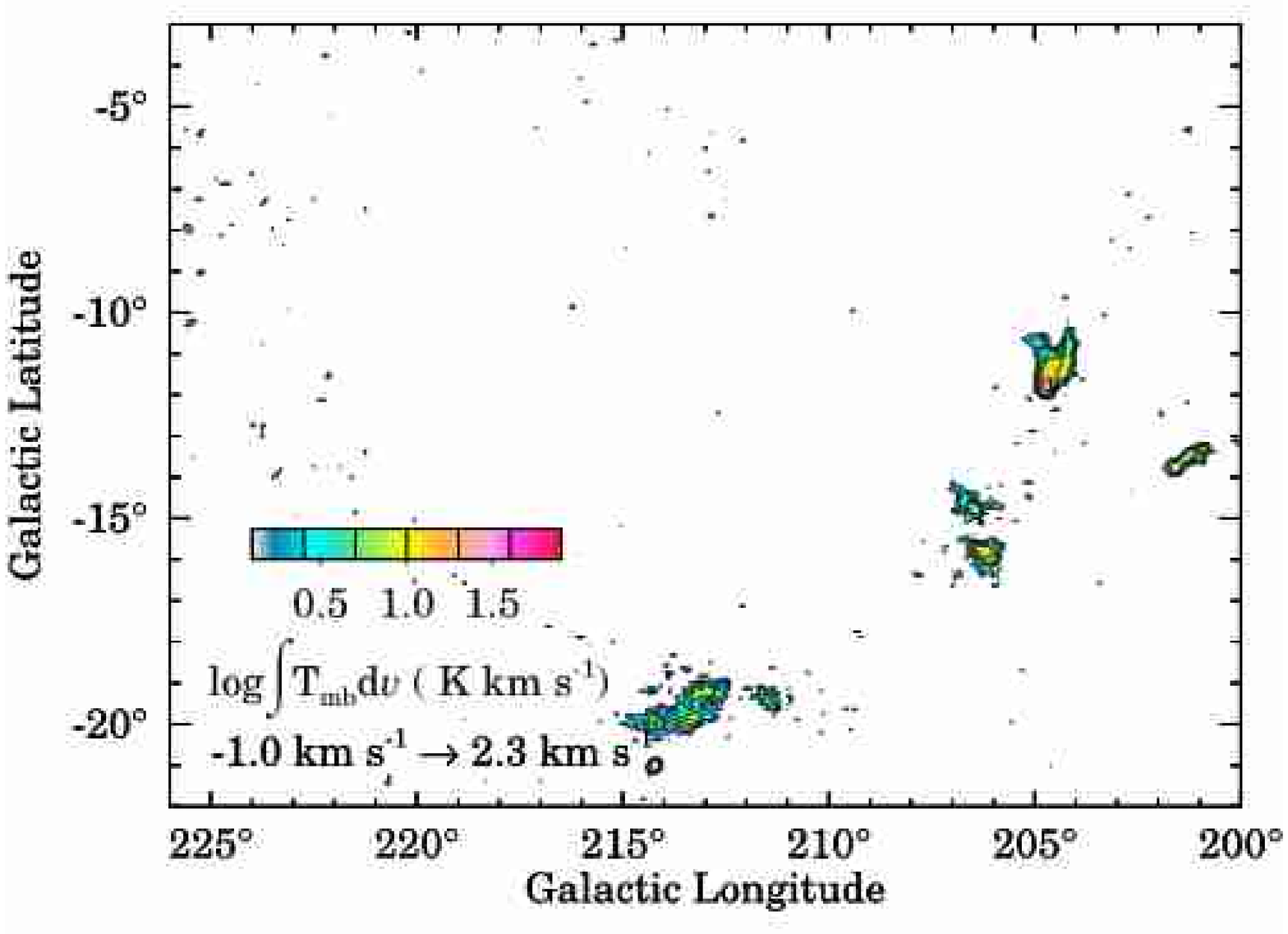}
\hfill
\vspace*{-0.4cm}
\epsfxsize=8.75cm
\epsfbox{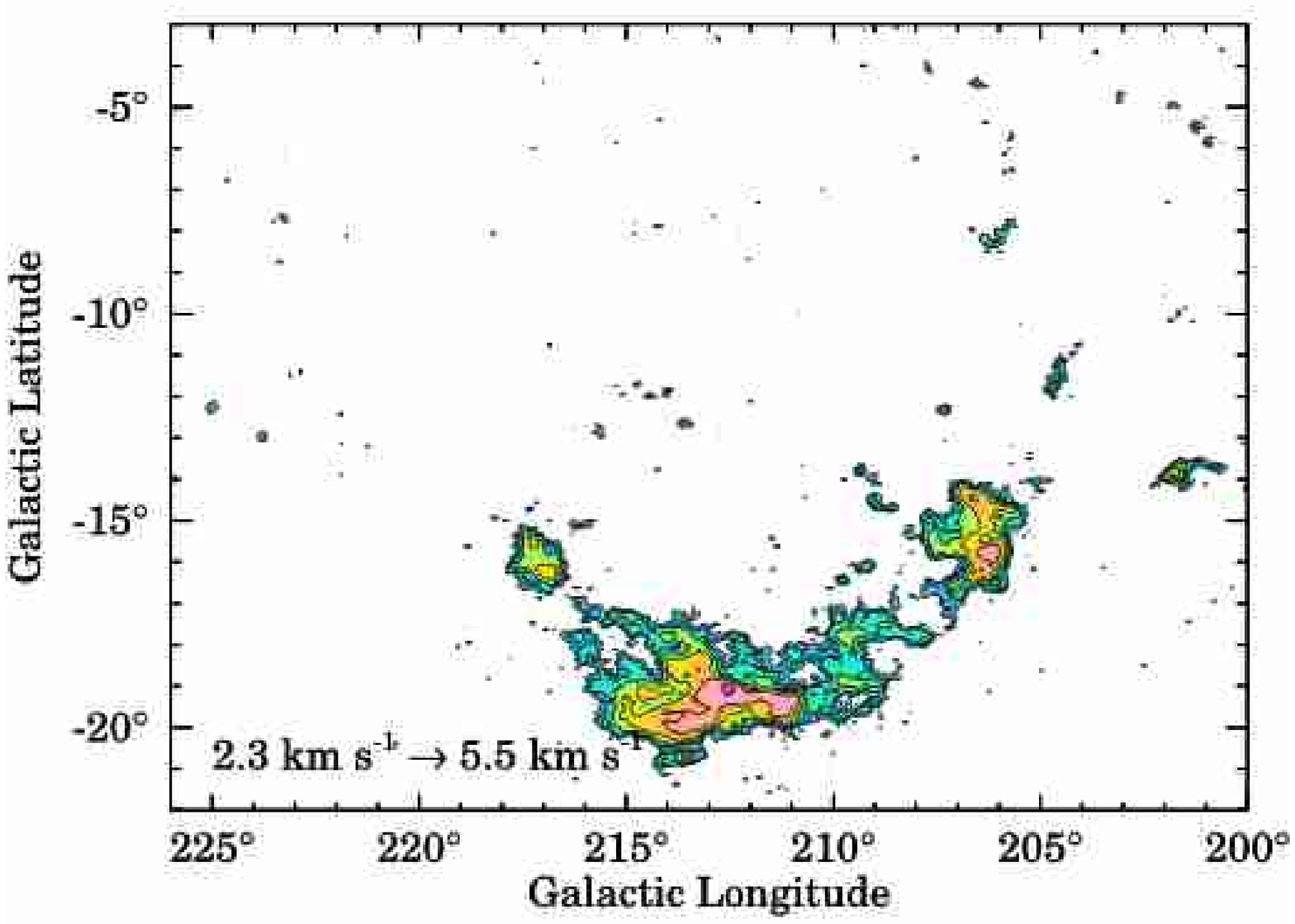}
\epsfxsize=8.75cm
\epsfbox{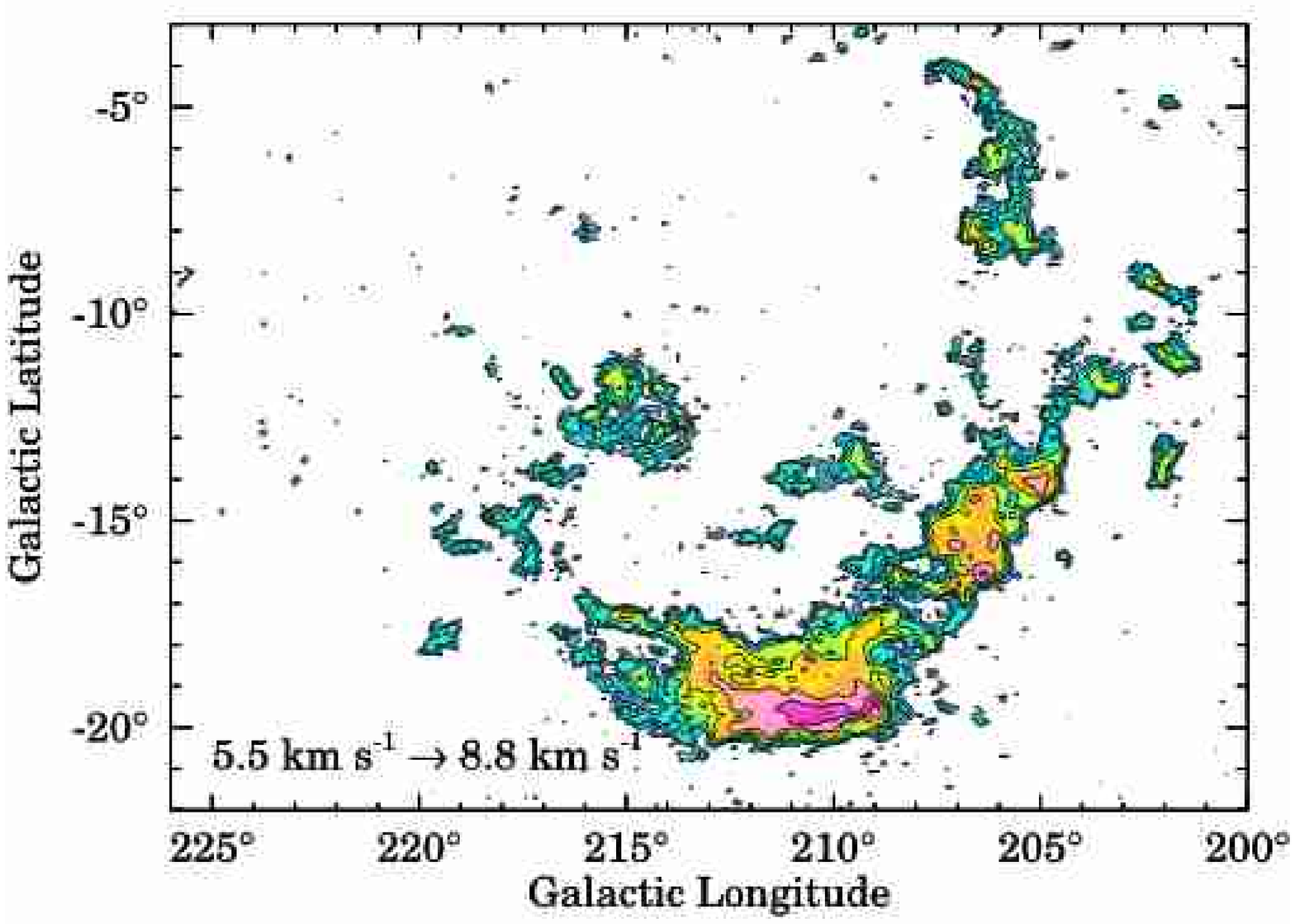}
\hfill
\vspace*{-0.4cm}
\epsfxsize=8.75cm
\epsfbox{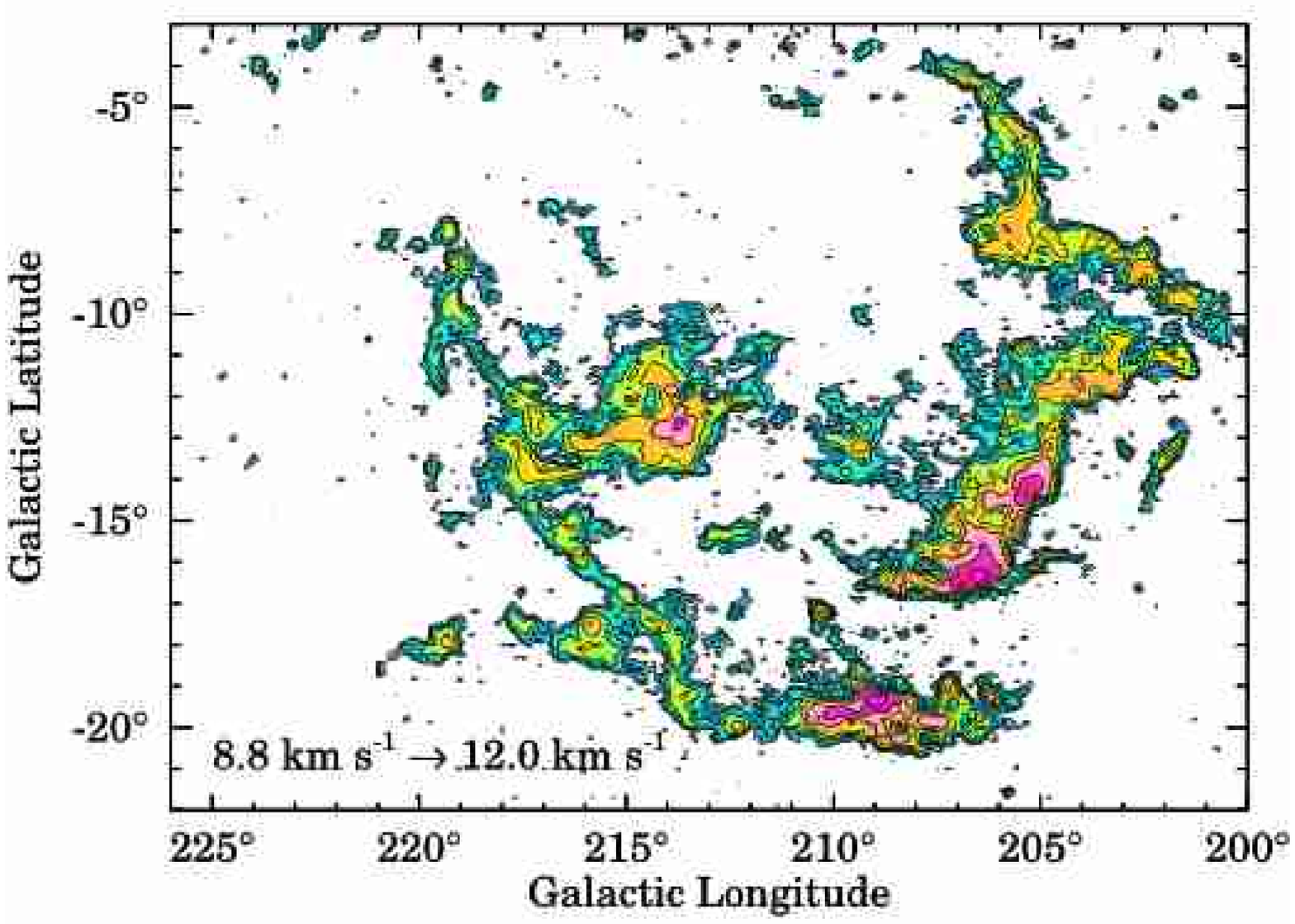}
\epsfxsize=8.75cm
\epsfbox{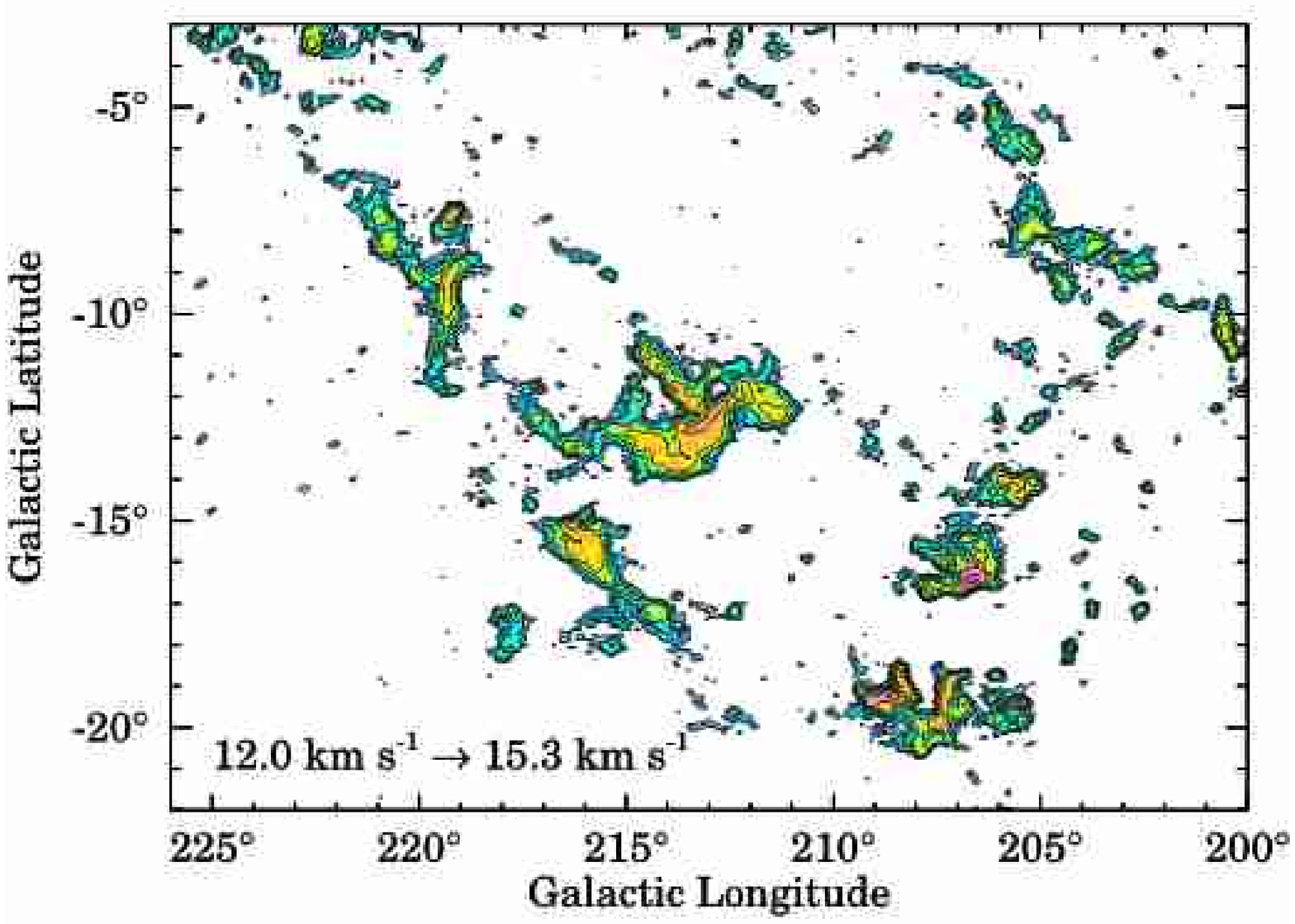}
\hfill
\epsfxsize=8.75cm
\epsfbox{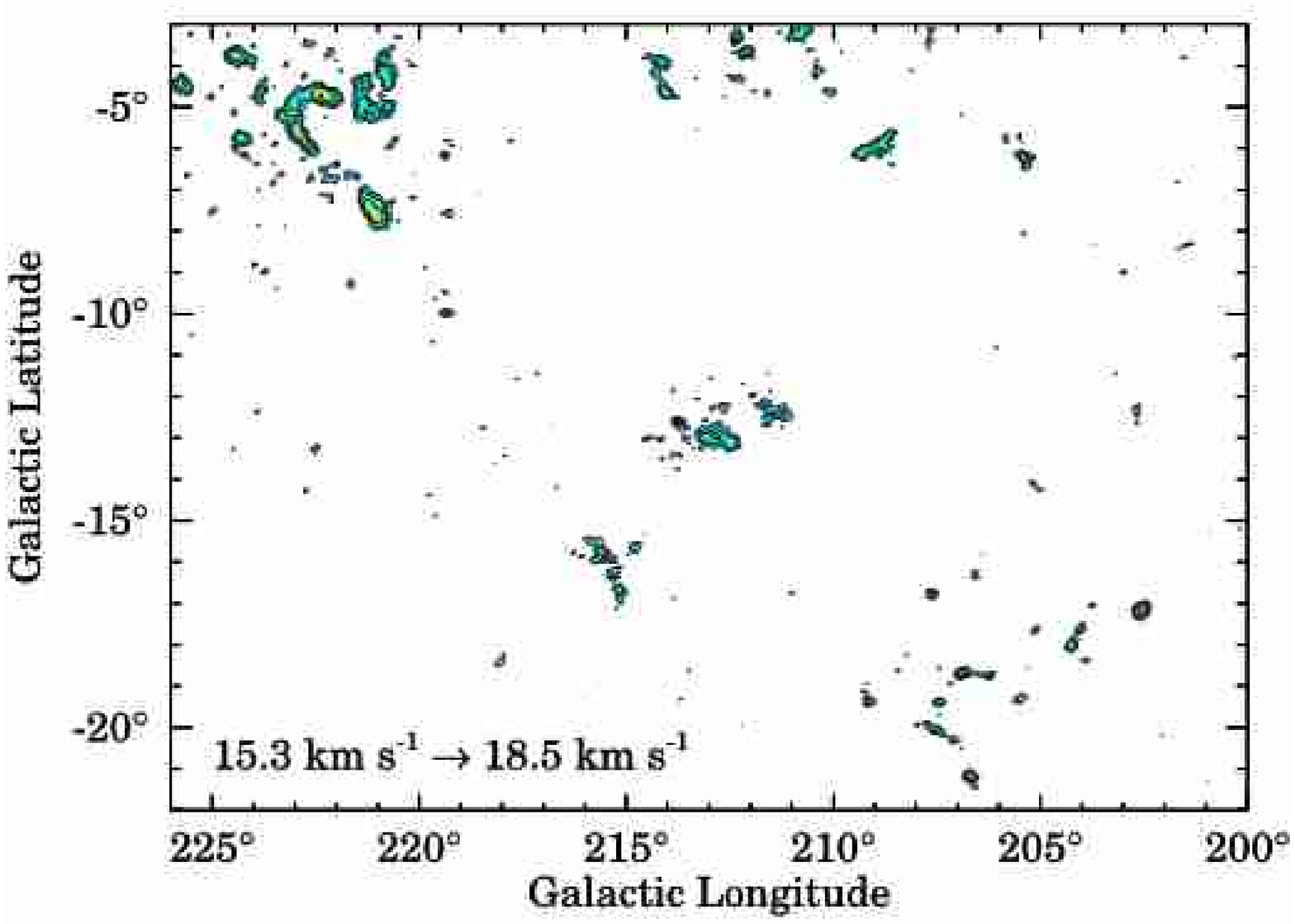}
\vspace*{-0.4cm}
\end{center}
\caption{Channels maps. Each map shows the integrated emission from five velocity channels (total width $\mathrm{3.25\,km\,s^{-1}}$) from $\mathrm{-1.0}$ to $\mathrm{18.5\,km\,s^{-1}}$. Log contour levels start at 0.1 $(3\sigma)$ and go up in steps of 0.3. }\label{orim3}
\end{figure*}

Measurements of the sky brightness as a function of elevation were
obtained by antenna tipping at least every six hours and more
frequently during periods of changeable weather, and were fitted to
the two layer atmospheric model of Kutner (1978)\nocite{kutner78} to
determine the temperature and opacity of atmospheric water vapour.
Further details on the receivers, observational technique and
calibration of the 1.2 m telescope can be found in Dame et
al. (1993).\nocite{Dame93} During the course of the observations total
system temperatures referred to above the atmosphere ranged from
$\sim$ 900\,K at the lowest elevations observed (30\degr) to $\sim$
500\,K at the highest ($\sim$ 60\degr).  Integration times were
automatically adjusted to achieve an rms noise temperature per channel
of $\sim$ 0.26~K.

Frequency switched observations were carried out over four consecutive
observing seasons from February 1999 to April 2002, using two 0\fdg125
grids that were offset by 0\fdg0625 in Galactic Longitude and
Latitude. The beam spacing of the final composite grid is 0\fdg09, and
the x and y axes of the grid are tilted at 45\degr\ with respect to
$l$ and $b$.  The frequency switched spectra were folded and
$5^{\mathrm{th}}$~order polynomial baselines fitted to the portion of
each spectrum that was found to be free of emission.  For each
observing session of $\sim 6$ hours a model of the CO emission feature
associated with the mesospheric layer of the Earth's atmosphere was
fitted to all spectra that were free of Galactic CO emission (Lang
1997).\nocite{Lang97} This model was then used to remove the
mesospheric CO line from all the spectra taken for that session.
Finally the oversampled observations were smoothed by convolution with
a two-dimensional Gaussian the size of the telescope's main beam in
order to lower the final rms noise to $\sim$ 0.14\,K.  The parameters
of the pre-smoothed observations are summarized in Table~\ref{param}.

\begin{figure}
\begin{center}
\leavevmode
\epsfxsize=8.5cm
\epsfbox{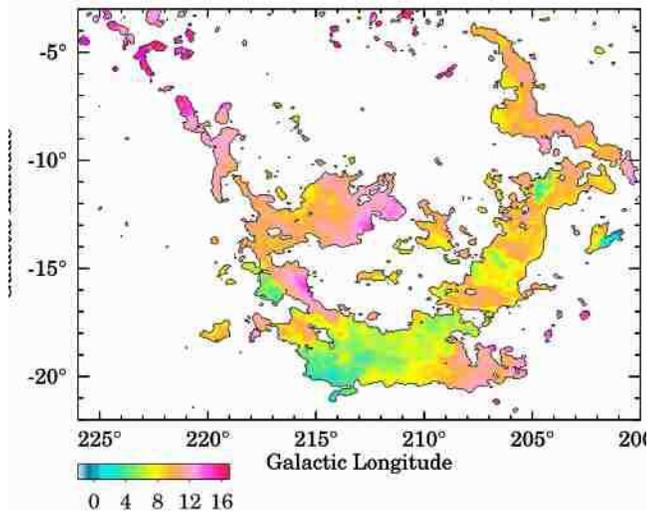}
\end{center}
\caption{Emission weighted mean velocity map of Orion-Monoceros. This figure clearly shows the velocity gradients along Orion~A and the Southern Filament and highlights regions with unusual kinematics such as the Scissors and NGC~2149.}\label{orim4}
\end{figure}

The CO observations are summarized in Fig.~\ref{orim2}, which shows
the spatial distribution of $\mathrm{W_{CO}}$
$\mathrm{(K\,km\,s^{-1})}$, the velocity-integrated line intensity.
The $v_\mathrm{LSR}$ integration interval of $\mathrm{-3}$ to
$\mathrm{19.5}$\,\kms\ was chosen after an examination of the channel
maps in order to include all the regions in the spectra where signal
was present.  The higher sensitivity per unit area of the new survey
is apparent in the amount of emission that is now detected from the
areas between the main molecular clouds.  Although weak, this emission
bridges the clouds spatially and without discontinuity in velocity,
indicating that none of the major clouds are entirely isolated.  The
velocity structure of the region is shown in the channel maps,
Fig.~\ref{orim3}, and the emission-weighted mean velocity map
Fig.~\ref{orim4}.  These two figures indicate that the majority of
the emission is confined to the relatively narrow velocity range $2
\rightarrow 15$\,\kms; although individual clouds such as Orion~A and
the Southern Filament have systematic velocity gradients, there is
none for the complex as a whole. The structure and kinematics of the
individual regions within the complex are discussed in detail in
\S\ref{sReg}.


\begin{table}
\caption{Parameters of the survey}
\label{param}
\centering
\vspace*{-0.3cm}
\begin{tabular}{llr}  \\ \hline \hline 
Number of spectra	&	52\,288 \\
Galactic longitude	&	200\degr$\rightarrow$226\degr  \\
Galactic latitude	&	-22\degr$\rightarrow$-3\degr \\
Angular resolution	&	8\farcm4  \\
Sampling interval	&	0\fdg088  \\
Velocity resolution	&	0.65 km\,s$^{-1}$  \\
Sensitivity		&	T$_{\mathrm{mb}}$= 0.26\,K rms \\
Frequency switching interval &	15 MHz \\ \hline 
\end{tabular}
\end{table}


\section{Molecular Masses and Cloud Distances}\label{sdist}

In spite of clear evidence of saturation in the denser cores of 
molecular clouds, the \co\ line remains the best overall tracer of 
mass in giant molecular clouds, since most of the mass is contained 
in the lower-density envelopes (Cambresy 1999).\nocite{Cambresy99}
Cloud masses in the Orion-Monoceros complex were estimated
using the standard assumption of a linear relationship between the
velocity integrated CO intensity, \wco, and the molecular hydrogen
column density, $N(\mathrm{H_2})$, taking
\[N(\mathrm{H_2})/\mathrm{W_{CO}}=(1.8\pm0.3) \times 10^{20}\,\mathrm{cm^{-2}\,K^{-1}\,km^{-1}\,s^{-1}}\]
as derived by Dame et al.(2001).  This yields
\[\mathrm{M_{CO}/M_{\sun}} = 1200\,\mathrm{S_{CO}}\,d^2_{\mathrm{kpc}}\]
where $d$ is the distance to the cloud in kpc, and $\mathrm{S_{CO}}$
is the CO emission integrated over velocity and the angular extent of
the cloud in K\,\kms\,arcdegrees$^2$.  A mean atomic weight of 1.36 is
assumed in order to account for helium and heavier elements (Allen
1973).

Historically the distances to Orion~A and B have been assumed to be
the same as the distance to the Orion nebula which has been estimated
photometrically as 435\,pc (Warren et al. 1977)\nocite{Warren77}, and
as 480$\pm$80\,pc via a study of the radial velocities and proper
motions of masers in the OMC\,1 core (Genzel et
al. 1981)\footnote{Mon~R2 is generally believed to be further away at
830$\pm$50\,pc (Racine 1968; Herbst \& Racine
1976)\nocite{Racine68}\nocite{Herbst76} and not directly related to
the other clouds in the complex.}.\nocite{Genzel81} Although these
estimates may be accurate, the implicit assumption that all the clouds
are at the same distance is unsatisfactory.  At 450\,pc the Orion~A
cloud would be over 100\,pc long, and although the region near to the
Orion nebula is likely to be at a distance close to the photometric
estimate, the other end of the cloud may be significantly nearer or
further away.


\newcolumntype{x}{D{,}{}{6}}
\newcolumntype{y}{D{,}{}{4}}
\newcolumntype{z}{D{.}{.}{3}}
\newcolumntype{a}{D{,}{}{8}}

\begin{table*}
\caption{Masses and distances for the molecular clouds in the Orion-Monoceros complex.}\label{dist}
\centering

\hsp

\begin{tabular}{lxyaz} \hline  \hline 

\multicolumn{1}{l}{Region}
& \multicolumn{1}{c}{Distance$^a$}
& \multicolumn{1}{c}{Previous}
& \multicolumn{1}{a}{\mathrm{Reference\ },}
& \multicolumn{1}{c}{Molecular} \\

\multicolumn{1}{l}{}
& \multicolumn{1}{c}{}
& \multicolumn{1}{c}{Estimate}
& \multicolumn{1}{l}{}
& \multicolumn{1}{c}{Mass\,$\times 10^3$} \\

\multicolumn{1}{l}{}
& \multicolumn{1}{c}{(pc)}
& \multicolumn{1}{c}{(pc)}
& \multicolumn{1}{l}{}
& \multicolumn{1}{c}{(\msun)}\\

\vspace*{-3ex}

\\ \hline 

Orion~A:\hspace*{20mm}\hfill	& 			& 		&							&   105.1		\\ 
\hspace{5mm}1\dotfill		& 521,^{+140}_{-91} 	& 		&   							&    12.3 		\\ 
\hspace{5mm}2\dotfill		& 465,^{+75}_{-57}	&  480,\pm80 	& \mathrm{Genzel\ et\ al.\ },\mathrm{(1981)}^b 		&   69.5		\\ 
\hspace{5mm}3\dotfill		& 412,^{+114}_{-74} 	&    		&							&   23.3		\\ 

\hts

NGC 2149\dotfill		& 425,^c  		& 	 	&							&   12.2 		\\ 

\hts

Orion~B:			&   			& 	 	&							&   82.3 		\\ 
\hspace{5mm}1 near\dotfill	& 422,^{+61}_{-48} 	&  400, 	& \mathrm{Anthony\!-\!Twarog\ },\mathrm{(1982)}  	&   11.9 		\\ 
\hspace{5mm}1 far\dotfill	& 514,^{+106}_{-75} 	&    		&							&   53.4 		\\ 
\hspace{5mm}2\dotfill		& 387,^{+112}_{-71} 	&    		&   							&   18.1		\\
\hspace{5mm}Scissors\dotfill	& \sim 150,^d 		&   		&   							&    0.340    		\\ 
\hspace{5mm}Orion East\dotfill	& 120,^{+20}_{-8} 	&   		&   							&    0.129		\\

\hts
	
Monoceros R2\dotfill		& 800,^{+300}_{-200}	&  830,\pm50	& \mathrm{Herbst\ \&\ Racine\ },\mathrm{(1976)}^e 	&   106.2		\\

\hts

Southern Filament:		&			& 	 	&							&   17.4		\\
\hspace{5mm}1 (Crossbones)\dotfill& 465,^{+169}_{-98}	&  	 	&							&   			\\
\hspace{5mm}2\dotfill		& 457,^{+150}_{-91}	& 	 	&							& 			\\

\hts
	
Northern Filament\dotfill	& 393,^{+65}_{-48}	&  	 	&							&   17.5		\\

\vspace*{-3.5ex}

\\ \hline

\end{tabular}

\vspace*{1mm}

\ssp

\parbox[t]{14cm}{$^a$ \parbox[t]{13.5cm}{$1\sigma$ uncertainties}}

\parbox[t]{14cm}{$^b$ \parbox[t]{13.5cm}{Also $\sim 435$\,pc Warren \& Hesser (1977)}}

\parbox[t]{14cm}{$^c$ \parbox[t]{13.5cm}{No background stars. Mean of distances to Orion~A (3) and Southern Filament (1) adopted.\vspace*{1ex}}}

\parbox[t]{14cm}{$^d$ \parbox[t]{13.5cm}{No foreground stars so distance estimated from background stars only.}}

\parbox[t]{14cm}{$^e$ \parbox[t]{13.5cm}{Also $\sim 950$\,pc Racine \& van den Bergh (1970)}}

\end{table*}
\nocite{Anthony-Twarog82}

In order to address this problem the parallaxes of stars observed with
the astrometric satellite \emph{Hipparcos} have been used to estimate
distances to the molecular clouds in Orion-Monoceros.  Individual
stars are identified as being in front or behind a particular
molecular cloud by comparing its colour with two theoretical models:
the colour of the star if it is reddened by dust associated with atomic
and molecular gas (i.e it is in the background of the cloud), and the
reddening associated with just atomic gas (i.e. it is in the
foreground of the cloud). The parallaxes of the stars that can be assigned
unambiguously to the foreground or the background of a molecular cloud are
then used to constrain the distance to that cloud.

In general \emph{Hipparcos} parallaxes for single stars are unreliable
for distances greater than $\sim 200$\,pc, but by using groups of
foreground and background stars and carefully modelling the
probability distribution function (pdf) for the cloud distance it is
possible to make robust distance estimates out to approximately
500\,pc. Furthermore, in some cases it has been possible to divide the
larger clouds into multiple sections and make separate distance
estimates for each part.  In these cases the velocity structure of the
complex was used as a guide when sub-dividing a cloud.

Table~\ref{dist} lists the distances and masses of the various clouds
in Orion-Monoceros estimated from \emph{Hipparcos} parallaxes. Most of
these distances agree, within the errors, with previous estimates made
using other techniques.  The distance determination technique is
described in more detail together with specific examples for Orion and
other local molecular clouds in Wilson et al. (2005).


\begin{figure*}
\begin{center}
\leavevmode
\epsfxsize=18cm
\epsfbox{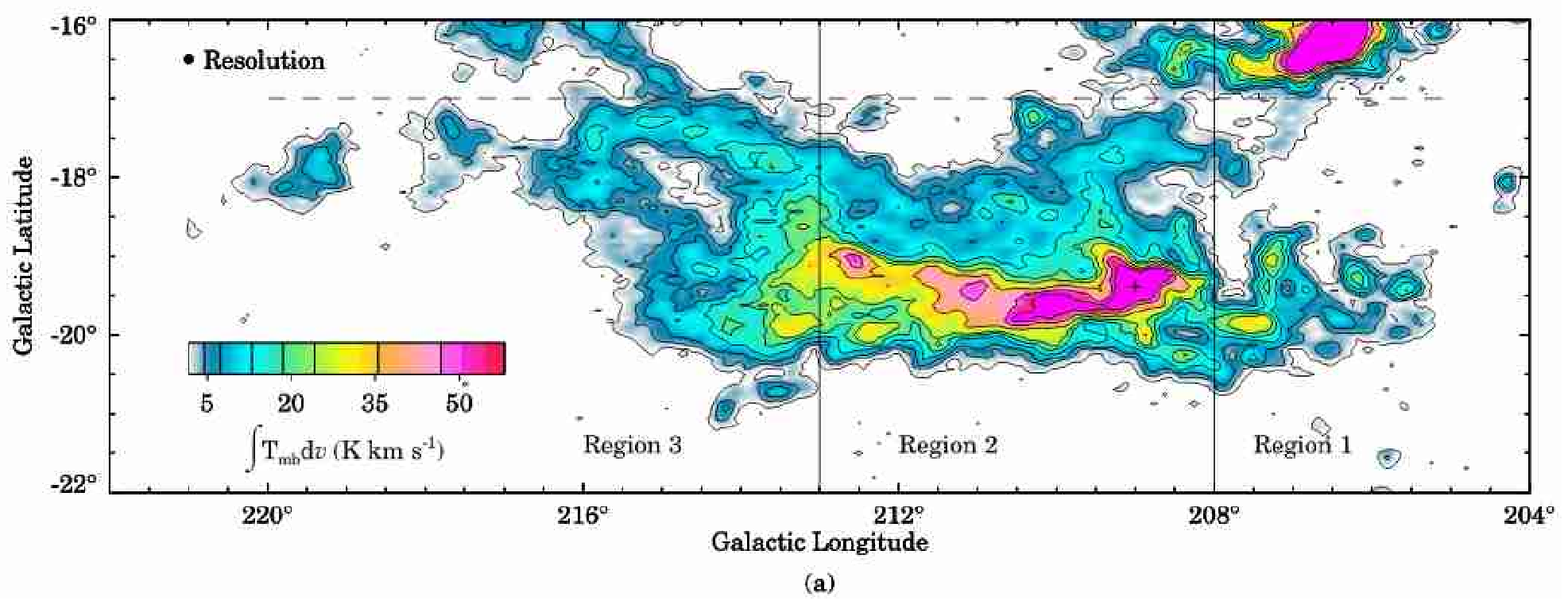}
\epsfxsize=18cm
\epsfbox{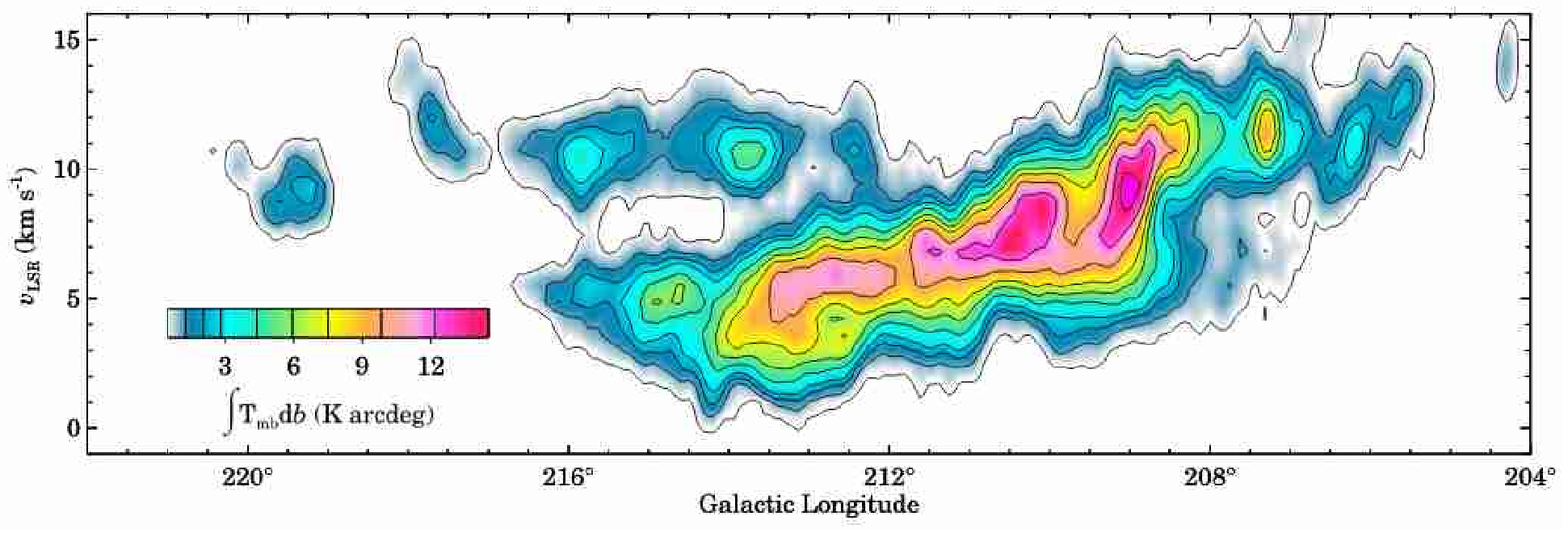}
\end{center}
\caption{\textbf{(a)} \wco\ map of Orion~A. The integration range is $0.0 < v_{\mathrm{LSR}} < 15.0$\,\kms.  Contours start at 1.68\,K\,\kms\ (3$\sigma$) and subsequent levels are at 8, 13, 23, 33, 43, 63, 83 and 103$\sigma$.  The dashed horizontal line marks the nominal boundary between Orion~A and Orion~B, the dotted vertical lines separate the three regions for which distances were estimated and the cross marks the position of the Orion Nebula. \textbf{(b)} Longitude-velocity map of Orion~A.  The integration range is $-22\degr\, < b < -17\degr$ and the contours start at 0.468\,K\,arcdegree (3$\sigma$) with subsequent levels at 3, 8, 13, 18, 28, 38, 48, 63, 78 and 93$\sigma$.}\label{orim7}
\end{figure*}

\section{Specific Regions}\label{sReg}

\subsection{Orion~A}

The Orion~A cloud, the largest in the Orion-Monoceros complex, is
located $\sim 19\degr$ below the Galactic plane at a distance of $\sim 450$\,pc
from the Sun.  It has a
surface area of $\sim 31.5$\,deg$^2$ ($\sim 2200\,\mathrm{pc^2}$) and
if cylindrical in shape, has a 
depth of $\sim 20$\,pc.  Fig.~\ref{orim7}~(a) shows the
velocity-integrated CO emission from Orion~A which is
somewhat cometary in appearance, particularly above the 24\,K\,\kms\
contour.  The apparent head of the comet is the compact knot of emission at
$l = 209\degr$ and $b = -19\fdg25$, and the ridge of dense gas that
points away from this peak (and also away from the centroid of the OB
association), becomes a wider lower density tail toward higher
longitudes.

The longitude-velocity map in Fig.~\ref{orim7}~(b) shows the
large-scale velocity gradient that runs along the length of the cloud.
First reported by Kutner et al.  (1977),\nocite{Kutner77} this
gradient has been confirmed by all subsequent large-scale observations
of the cloud. Its origin has been attributed to both rotation (Kutner
et al. 1977; Maddalena et
al. 1986)\nocite{Kutner77}\nocite{Maddalena86} and to large scale
expansion driven by the stellar winds of the Ori OB1 association
(Bally 1987).\nocite{Bally87} In the first case the rotation is about
an axis perpendicular to the Galactic plane and opposed to the
rotation of the Galaxy.  Heyer et al. (1992) argue
\nocite{Heyer92}that this model is unlikely to be correct as the
velocity-gradient (as shown by their data) is not sufficiently
continuous over the length of the major axis to be consistent with the
cloud's rotational motion.  Specifically, there exist sharp velocity
shifts of $\sim 1$\,\kms\ between adjacent spectra at many locations.
Bally et al (1987)\nocite{Bally87} and Heyer et al. (1992)
\nocite{Heyer92}also identify many regions within Orion~A that exhibit
very sharp emission gradients and velocity shifts which they attribute
to expanding shells driven by HII regions excited by young massive
stars.  However, the velocity gradient as shown in
Fig.~\ref{orim7}~(b) extends over a distance of $\sim 100$\,pc and
it is unlikely that such a coherent structure is the result of purely
internal processes. It is possible, however, that the gradient was
formed by the passage through the cloud of a much larger shell,
similar to those driven by the Ori\,OB association.
It is noted that it is not possible to see the huge Orion~A outflow in the
longitude-velocity (Fig.~\ref{orim7}~(b)) as the data has been integrated 
over 5\degr.  However, the Orion~A outflow can be seen clearly in narrower
position-velocity slices (see \S\ref{large}). 

Distances to Orion~A were determined for each of three sections;
delineated by the vertical dotted lines on Fig.~\ref{orim7}~(a) at
208\degr\ and 213\degr, using the \emph{Hipparcos} constraining star
method described in \S\ref{sdist}.  The distances were estimated to be
520\,pc, 465\,pc and 410\,pc for Regions 1, 2 and 3 respectively.
Although adjacent distance estimates are roughly equal
(c.f. Table~\ref{dist}), they imply that the cloud is oriented in a
direction consistent with the rotation model of Kutner (1977) and
Maddelena (1986).  However, this model is unable to explain the
velocity jump between Regions 1 and 2.  A more likely explanation for
the different distances is suggested by the locations of the stars of
the OB\,1b subgroup (c.f. Fig.~\ref{orim1}) which are believed to be
located at a distance of $\sim 360$\,pc (Brown et
al. 1994)\nocite{Brown94} in front of the low-longitude end of the
cloud.  A model where Region~1 is accelerated away from the Sun by the
energy released from the massive stars of OB\,1b is entirely
consistent with the kinematics of the cloud. Furthermore there are
several small clouds (for example near $l=202\fdg5, b=-17\degr$) located
toward the OB\,1b association that also have high velocities
consistent with this model.

If the low-longitude end of Orion~A has been accelerated away from the
Sun by the stellar winds of the OB\,1b subgroup then we would expect
to see a corresponding lateral acceleration of the gas away from the
centre of the subgroup.  The $-1.0 \rightarrow 8.8$\,\kms\ channel
maps (Fig.~\ref{orim3}) show that the area around $l \approx 205\degr, b \approx -19\degr$ (the part of the cloud closest to the OB\,1b
subgroup) has been cleared
of molecular gas at low velocities ($\leq 5$\,\kms).  In this velocity
range the gas in Orion~A, and on the low-latitude side of Orion~B, may
also have been compressed laterally, a process that has given rise to the
pronounced step in the Orion~A longitude-velocity map at
$\sim$208\fdg5.  The velocity step is coincident with both the peak CO
emission and the position of what is presently the most active site of
star-formation within the complex, the Orion nebula
\nocite{Hillenbrand97}(Hillenbrand 1997). The fact that this region of
high density gas (at $\sim 465$\,pc) and the young stars of the Orion
Nebula (at $\sim 435$\,pc; Warren \& Hesser 1977)\nocite{Warren77}
appear to be at similar distances is not entirely unexpected. It is
difficult to avoid the conclusion that the OB\,1b subgroup compressed
the gas in the proto-Orion~A cloud which triggered the formation of
the OB\,1c and OB\,1d associations.

Moving along Orion~A away from the centre of the OB1c subgroup and
toward higher longitudes, the gas becomes more diffuse and there is a
marked decrease in star-formation.  Region 3 of the
Orion~A cloud, the area between $l \approx 213\degr$\ and $l \approx
217\degr$, contains two components that are distinct both in position
and velocity. Maddalena et al. (1986) suggested that the second,
higher velocity component belonged to a separate group of clouds
associated with NGC~2149 that is not directly related to Orion A.  It
will be shown in section~\ref{ring} that the NGC~2149 clouds in fact
form a coherent structure with the high longitude end of the Orion~A
cloud and part of the Southern Filament.

\begin{figure*}
\begin{minipage}[b]{18cm}
\begin{center}
\leavevmode
\epsfxsize=10.8cm
\epsfbox{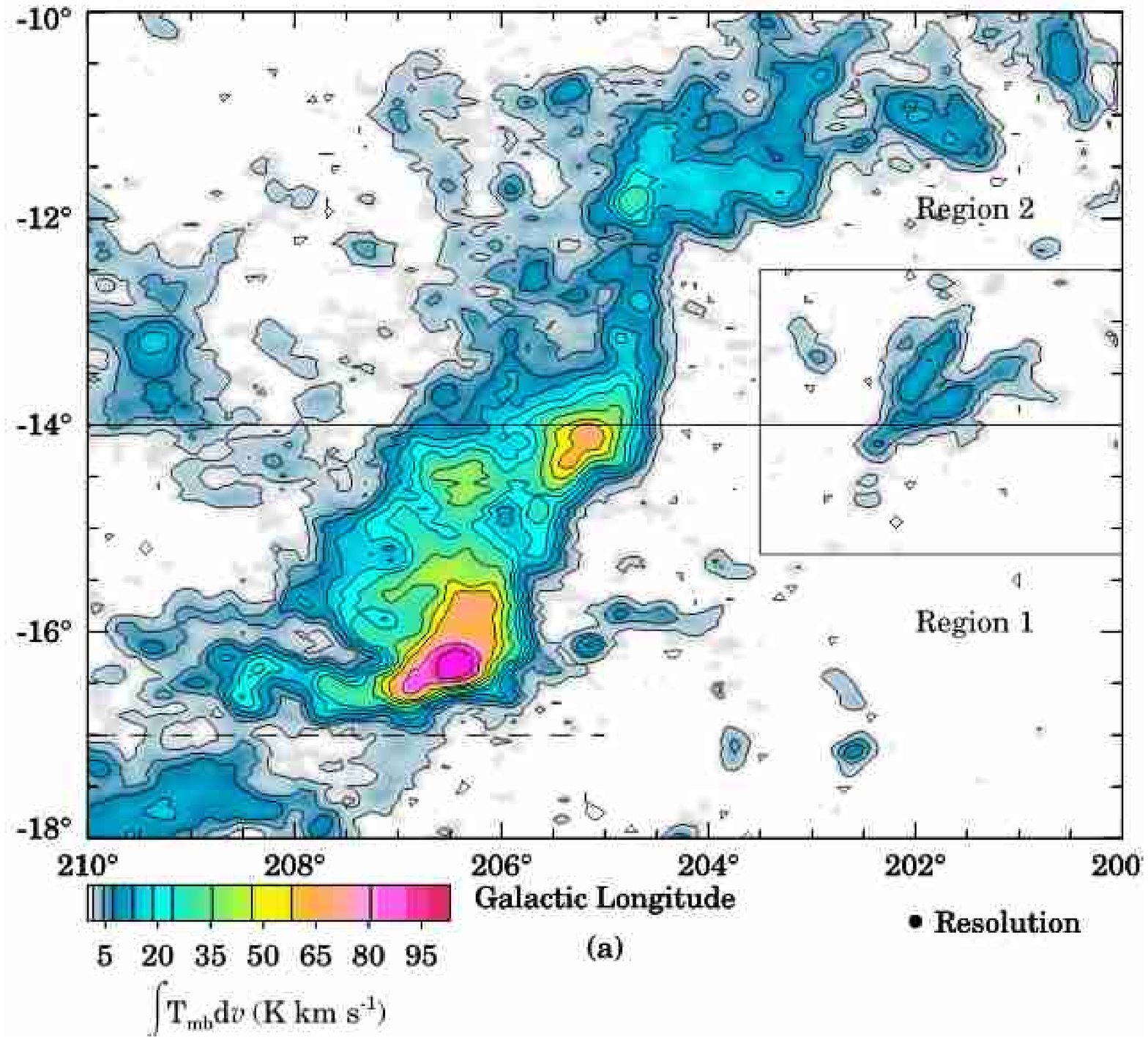}
\hfill
\epsfxsize=6.8cm
\epsfbox{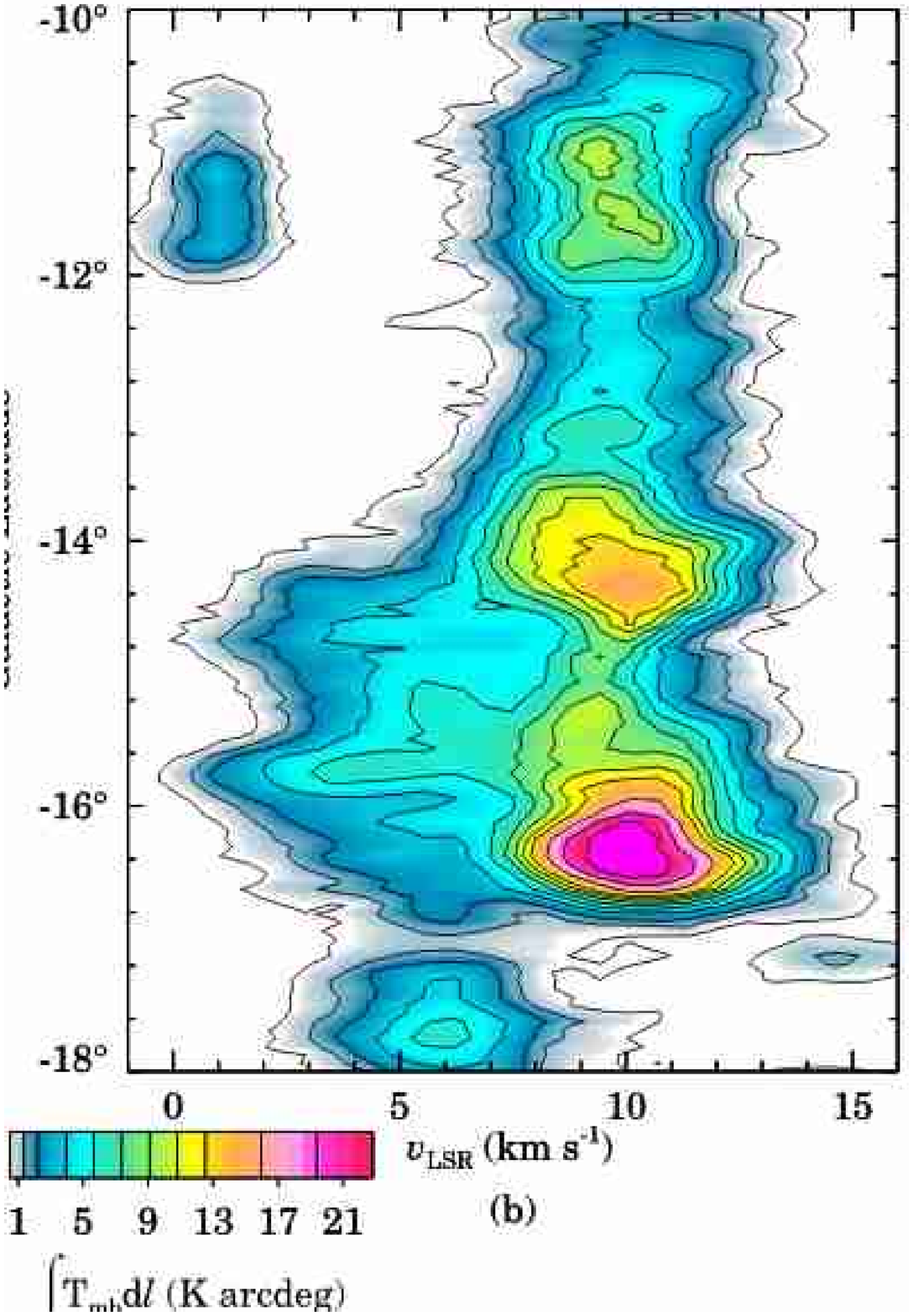}
\end{center}
\end{minipage}
\caption{\textbf{(a)} \wco\ map of Orion~B.  The integration range is $0 < v_{\mathrm{LSR}} < 15$\,\kms.  Contours start at 1.68\,K\,\kms\ (3$\sigma$) and subsequent levels are at 8, 13, 23, 33, 43, 63, 83, 103, 143 and 183$\sigma$. The horizontal dashed line at $b = -17\degr$\ marks the nominal boundary between the Orion~A and Orion~B clouds.  The horizontal dotted line at $b = -14\degr$\ separates the clouds into two regions for which distances were estimated.  Emission from ``the Scissors'', marked by the dashed box was not included in the velocity-latitude map.  \textbf{(b)} Velocity-latitude map of Orion~B.  The integration range is $100\degr\, < l < 210\degr$ and the contours start at 0.514\,K\,arcdeg (3$\sigma$) with subsequent levels at 8, 13, 23, 33, 43, 53, 63, 73, 93, 113 and 133$\sigma$.}\label{orim9}
\end{figure*}

\subsection{Orion~B}

Orion~B is the second largest cloud in the Ori-Mon complex, at least
in apparent size.  The cloud subtends an area of $\sim 25.7$\,deg$^2$,
which at $450$\,pc is $ 1500\,\mathrm{pc^2}$, and it has a mass of $0.8
\times 10^5$\,\msun.  The CO emission from Orion~B is summarized in
Fig.~\ref{orim9}, which shows (a) the $\mathrm{W_{CO}}$ map and (b)
the latitude-velocity map.  The dashed line at $b = -17\degr$\ on the
\wco\ map marks the nominal boundary between the Orion~B and Orion~A
clouds.  Emission from the ``Scissors'' (discussed below) was not
included in the latitude-velocity map.

The emission from Orion~B peaks at $l = 206\fdg5, b = -16\fdg5$,
toward the HII region NGC~2024 and the reflection nebula NGC~2023.  A
ridge of high density gas extends from the CO peak along the very well defined
low-latitude edge of the cloud.
The low-longitude side of the cloud, the side closest to the centres
of the OB\,1a and OB\,1b subgroups, also shows a sharp edge in CO
emission.  These sharp edges suggest that the boundary between the
molecular and the atomic gas is seen close to edge on.  By contrast
there is a network of increasingly diffuse ridges and filaments that
trail off toward the opposite side of the cloud, in directions that
point away from the centre of the OB association.

It can be seen from the velocity-latitude plot in Fig.~\ref{orim9}
(b) that the gas in Orion~B does not possess a systematic velocity
gradient and that the cloud as a whole is centered at $v \approx
10$\,\kms.  However, the majority of spectra from the region below $b
= -14\degr$\ contain additional emission at a lower velocity ($v
\approx 7$\,\kms).  This excess can also be seen in the low-velocity
channel maps (Fig.~\ref{orim3}).  Many of the CO lines from this
region are either double peaked or unusually wide and lopsided with
the excess at low velocities.  Toward higher Galactic latitudes,
around $l = 204\fdg5$, $b = -11\fdg5$, a second cloud is superimposed
upon Orion~B.  Associated with the dark nebulae L\,1621 and L\,1622
(Lynds 1962)\nocite{Lynds62} and designated Orion East by Maddalena et
al. (1986), this cloud is at a velocity of $\sim 1$\,\kms\ and may
have little connection to the main Orion~B cloud, though it seems
kinematically related to the Orion-Eridanus superbubble. Orion East is
discussed in the next section.

Orion~B was divided into two sections, and distances estimated for
each.  The area located below $b = -14\degr$ which contains the CO
peaks associated with NGC~2023, NGC~2024, NGC~2068 and NGC~2071 was
designated Region~1.  The region of more diffuse gas above $b =
-14\degr$ was designated Region~2.  When the foreground--background
assignment was made for Region~1 there were three stars whose
extinction was four times below the value that would be expected if they
were located behind the cloud, but $\sim3$ times above the extinction
predicted by HI alone.  The dust associated with the molecular gas in
the low velocity tail of Orion~B can probably account for the
extinction in excess of that expected from HI, suggesting that the
three slightly reddened stars are located within the low velocity tail
of Orion~B which extends away from the main cloud toward the Sun. Two
distances to Region 1 were estimated, a near distance of $\sim
420$\,pc and a far distance of $\sim 520$\,pc.  The distance to Region
2 was found to be $\sim 390$\,pc, though this estimate is rather
uncertain as only a small number of suitable \emph{Hipparcos} stars
are found toward this part of Orion~B.

A kinematic model where the stellar winds from the Ori~OB\,1b subgroup
have reached Orion~B at its low-latitude and low-longitude corner is
consistent with the observed distances.  The gas at
the interface has been compressed laterally and accelerated away from
the Sun.  This gas is located at the far distance.  However, a small
fraction of the gas has been accelerated toward the Sun.  This gas,
which does not appear to be fragmented (c.f the $2.3 \rightarrow
5.5$\,\kms\ channel map - Fig.~\ref{orim3}), is estimated to have a
mass of $1.2 \times 10^4$\,\msun, or about 15\% of the total mass of
the cloud.

\subsection{Orion East}

\begin{figure}
\epsfxsize=8.5cm
\epsfbox{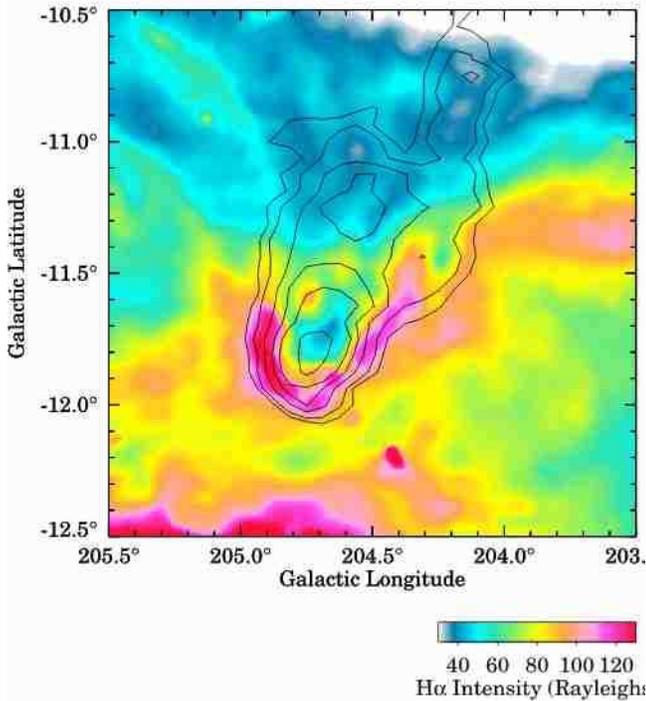}
\caption{Ionized gas around Orion East (L\,1622). The colour field represents the smoothed continuum subtracted H$\alpha$ emission mapped by the Southern H$\alpha$ Sky Survey (SHASSA, Gaustad et al. 2001). Velocity-integrated CO emission from the globules is represented by the superimposed contours.}\label{orim11}
\end{figure}

The contours in Fig.~\ref{orim11} are those of velocity-integrated
CO in the cometary Orion East cloud, located toward the high latitude
part of Orion~B. The H$\alpha$ emission from the region, shown in
false colour in Fig.~\ref{orim11}, indicates that the ionized gas
is streaming around the cloud. The pressure of this gas may have
triggered the recent star-formation that is suggested by the presence
of at least five T\,Tauri stars (Herbig \& Rao 1972; Cohen \& Kuhi
1979)\nocite{Herbig72}\nocite{Cohen79} in the head of Orion East.

Orion East may be at a distance of only 120\,pc --
significantly closer than Orion~B and the only previous
distance estimate of $500\pm 140$\,pc by Herbst
(1982).\nocite{Herbst82} The new estimate is based upon the parallaxes
of only three \emph{Hipparcos} stars, but since they all had
small relative errors and could be assigned unambiguously, the new
estimate is probably robust.  In the kinematic model of the
Orion-Eridanus bubble of Bally et al. (1998), Orion East has been
accelerated toward the Sun and is on the near side of the
bubble.  The new distance estimate is in agreement with this model.
It also agrees with estimates concerning the distance to
the Orion-Eridanus bubble based upon X-ray intensity maps (Guo et
al. 1995) \nocite{Guo95}which indicate that the near side of the
bubble is at $159\pm 16$\,pc and the far side is at $226
\pm 24$\,pc (Guo et al. 1995) or $\sim 400$\,pc (Reynolds \& Ogden
1979).\nocite{Reynolds79}

Orion East is also noteworthy because it lies beyond the projected
edge of the H$\alpha$ emission associated with Barnard's Loop and the
Orion-Eridanus bubble. Bally et al. (1998) argued that the
Orion-Eridanus ionization boundary had receded, and that Barnard's
Loop (and its faint H$\alpha$ extensions into Eridanus) delineate the
current ionization boundary of the Orion-Eridanus bubble, but in the
past the influence of the OB association must have extended beyond
this boundary in order to shape Orion East and the L\,1617 globules
into their present cometary form.  However, Fig.~\ref{orim11} shows that
ionizing gas continues to stream around Orion East, and so a more
likely explanation is that Barnard's Loop traces the ionization front
of the near side of the Orion-Eridanus bubble, which is much closer to
the Sun than the Orion-Monoceros molecular clouds.

\subsection{VDB 49: the Scissors}

\begin{figure}
\begin{center}
\leavevmode
\epsfxsize=8.5cm
\epsfbox{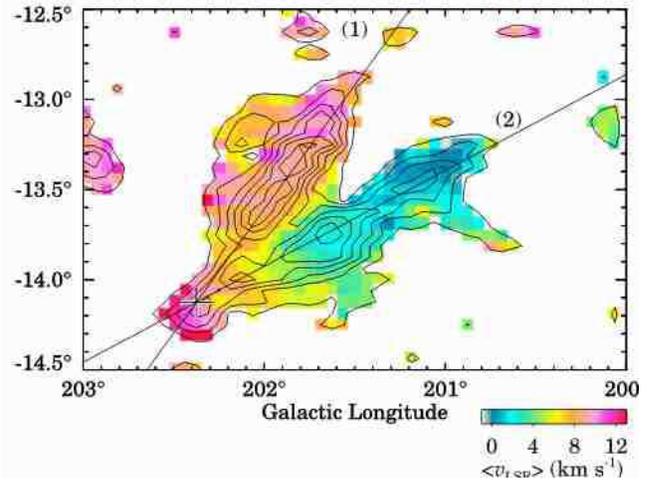}
\end{center}
\caption{The Scissors. The colourscale represents the emission weighted mean velocity and the contours the velocity-integrated CO intensity. The integration range is $-1 < v_{\mathrm{LSR}} < 16$\,\kms\ and the contours are at  1.8, 3.6, 5.4, 7.2, 9.0, 10.8 and 14.4\,K\,\kms. The two solid lines mark the positions of the position-velocity slices in Fig.~\ref{orim13}.  The cross marks the position of zero offset in these slices.}\label{orim12}
\end{figure}

\begin{figure}
\begin{center}
\leavevmode
\epsfxsize=4.2cm
\epsfbox{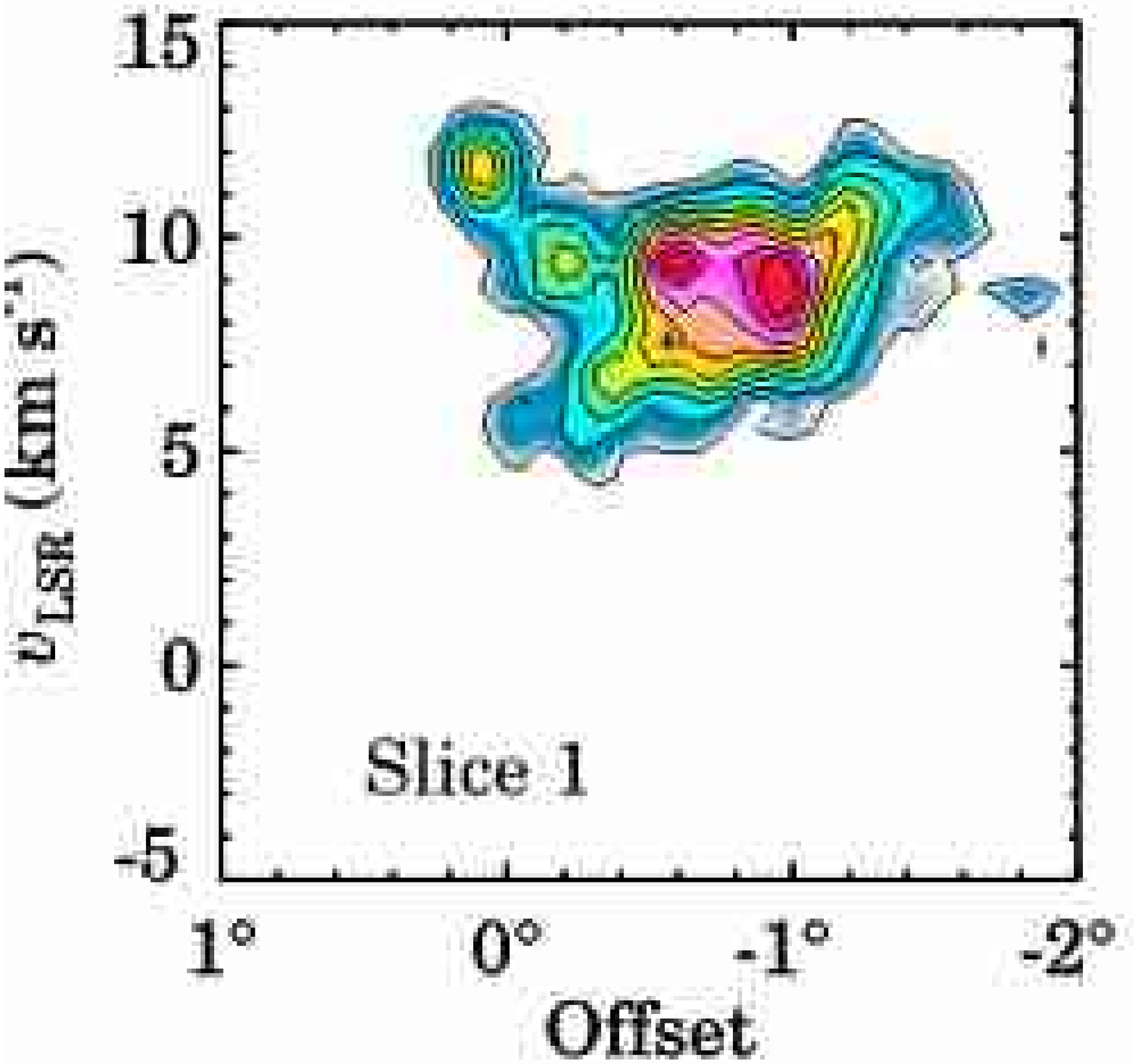}
\epsfxsize=4.2cm
\epsfbox{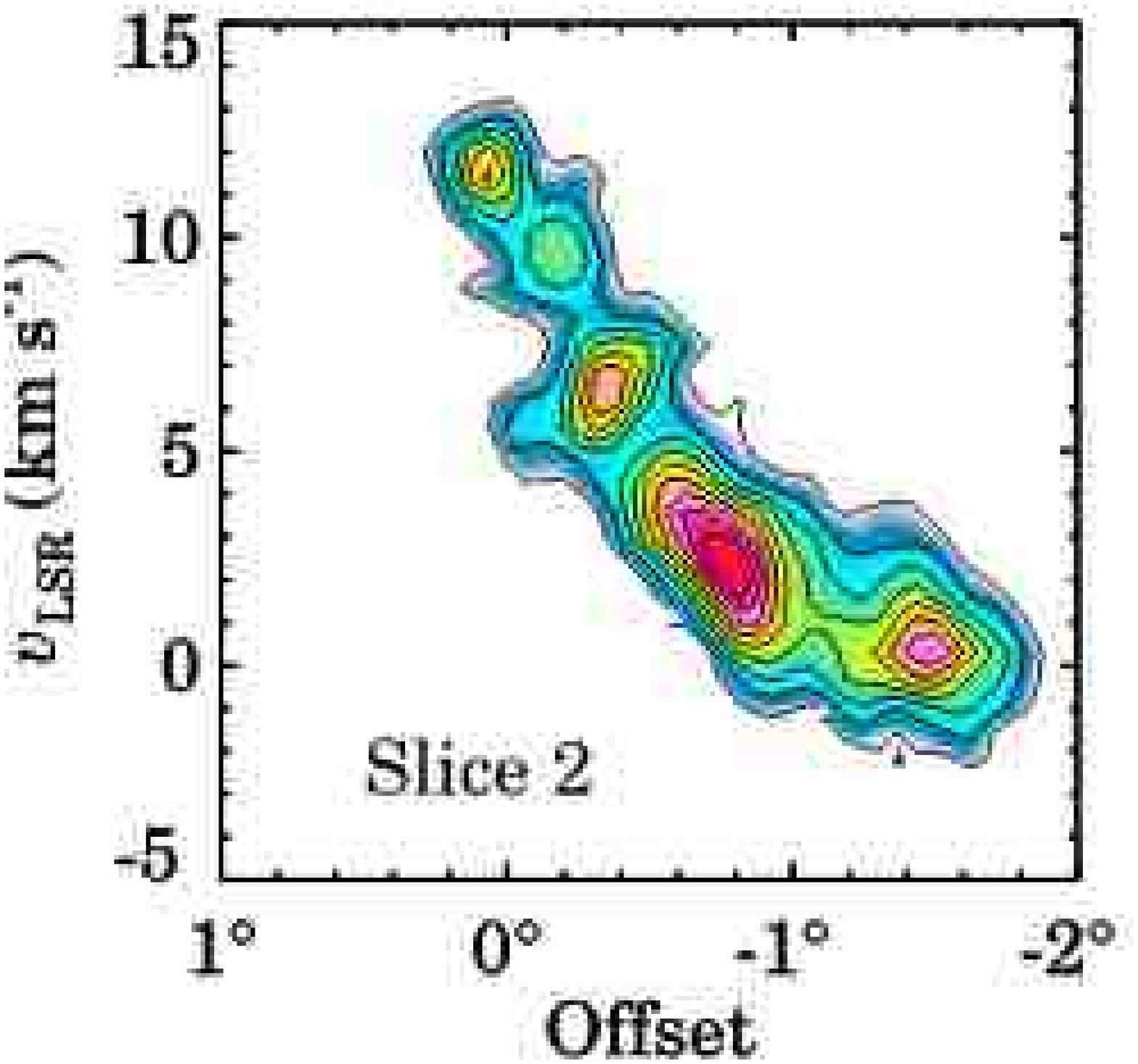}
\end{center}
\caption{Position-velocity slices through the Scissors.  The orientation of the slices and the zero offset position are shown in Fig.~\ref{orim12}. Each slice is integrated over a width of 0\fdg375 wide.  The contours in both plots start at 0.1\,K\,arcdeg with subsequent steps at each additional 0.1\,K\,arcdeg.}\label{orim13}
\end{figure}

At $l = 202\degr$, $b = -14\degr$
is an unusual molecular cloud that we have dubbed the Scissors.
The cloud covers an area of $\sim 2.2$\,deg$^2$, which at a distance of
$\sim 150$\,pc is equivalent to $\sim 15\,\mathrm{pc^2}$, and it has a
mass of $\sim 140$\,\msun.  The distinctive two limb
structure and unusual kinematics of the cloud are summarized in
Fig.~\ref{orim12}, which shows the weighted mean velocity field of the
cloud as a colour field with contours representing the
velocity-integrated CO emission superimposed.  The solid lines mark
the positions and orientations of the two position-velocity slices,
one along each limb, that are presented in Fig.~\ref{orim13}.

The Scissors appear to be a large cometary
cloud with a single head that points toward the centre of the OB\,1b
subgroup and a forked tail that extends toward the Galactic plane.
However, the two branches of the tail have very different kinematics 
as shown by the position-velocity slices, Fig.~\ref{orim13}.
The lower limb (slice 2) has a
steep, smooth velocity gradient along its entire length.  This
gradient is largely absent in the upper limb (slice 1). The
apparent velocity discontinuity at the position where the two limbs
intersect suggests that the Scissors are two separate
clouds along the same line of sight.  However,
the probability of the chance alignment of two similarly shaped and 
oriented clouds 14\degr\ from the Galactic plane is small.

It is not possible to determine the distance to either branch via the
constraining star method as there are no \emph{Hipparcos} stars that
can be reliably assigned to the foreground of the Scissors.  However,
there are two stars, one for each branch, that can be assigned to the
background and these indicate that neither branch of the Scissors is
further away than $\sim 300$\,pc.  We tentatively adopt a distance of
$\sim 150$\,pc to both. It is noted that the Scissors does not display
a strong correlation between H$\alpha$ emission and \wco\ comparable
to that of the nearby cloud Orion-East, which has a similar position
close to Barnard's Loop and a more secure distance.

\begin{figure}
\begin{center}
\leavevmode
\begin{minipage}[b]{8.5cm}
\epsfxsize=8.5cm
\epsfbox{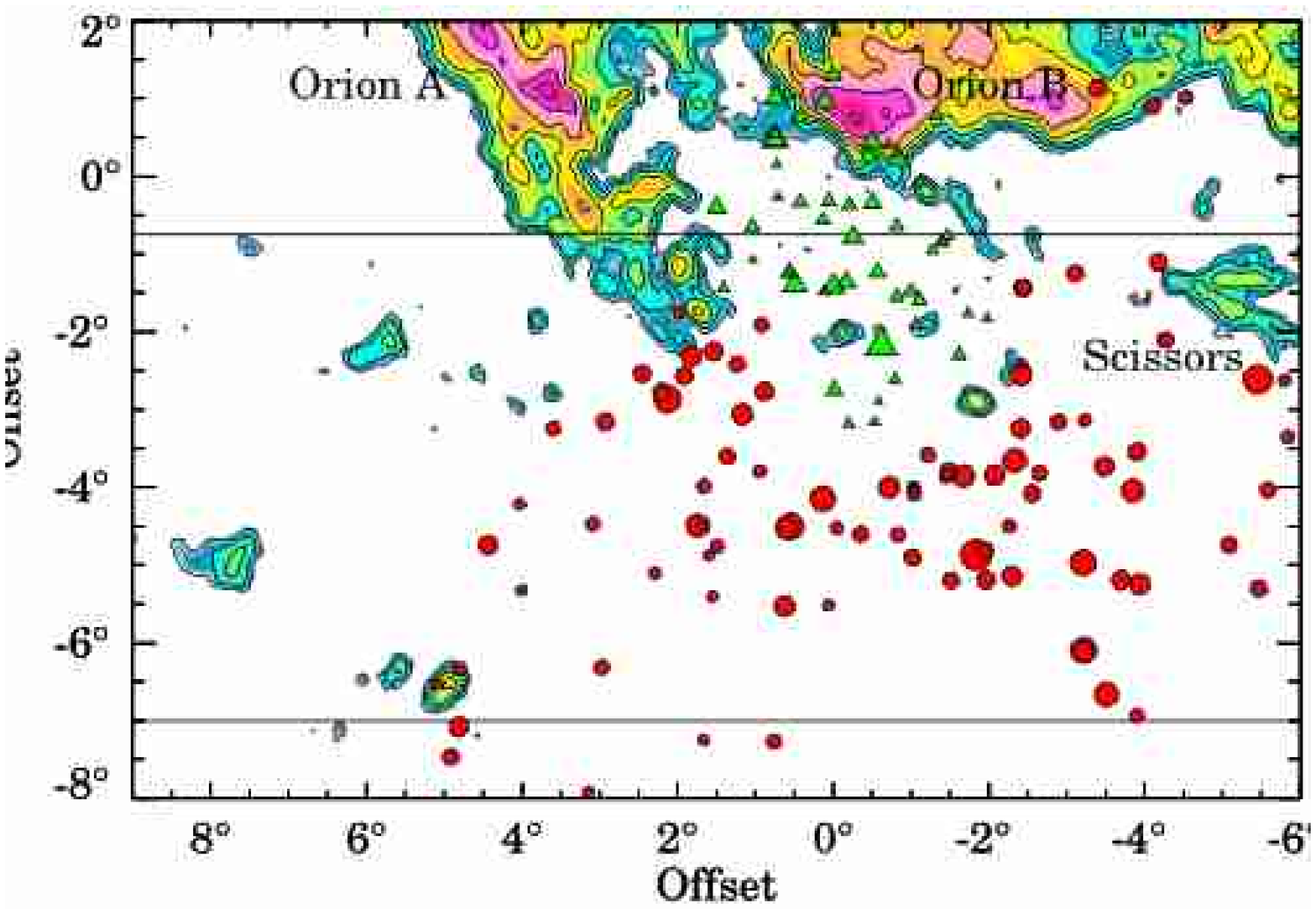}
\end{minipage}
\vspace*{0.45cm}
\vfill
\begin{minipage}[b]{8.5cm}
\epsfxsize=8.5cm
\epsfbox{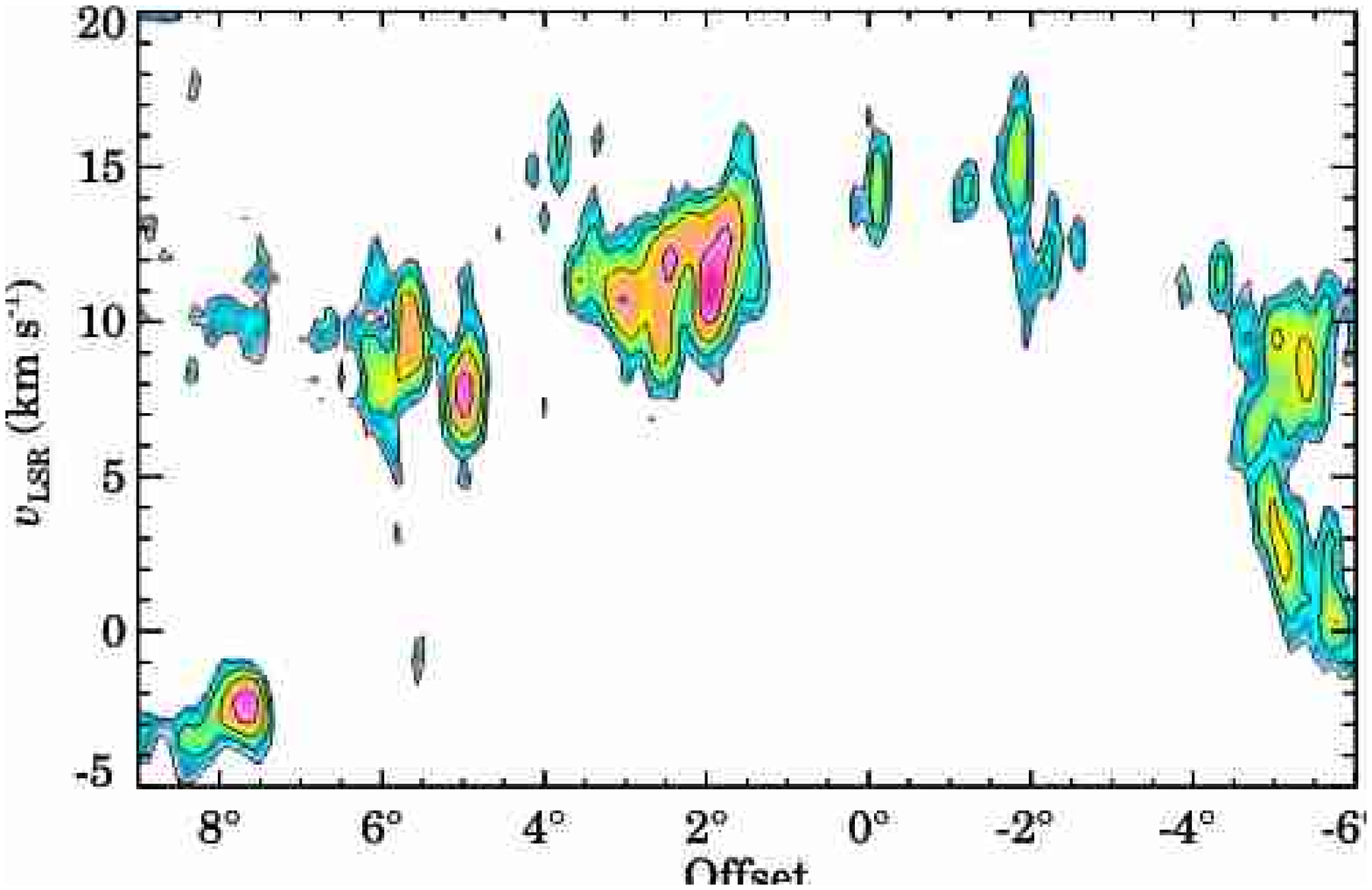}
\end{minipage}
\end{center}
\caption{Position-velocity slice through Ori~OB\,1a. \textbf{(a)} \wco\ map rotated 55\degr\ clockwise about the position $l = 206\degr$, $b = -17\degr$. The positions of O and B stars from the Ori\,1a and Ori\,1b subgroups are plotted with the same symbols as Fig.~\ref{orim1}.  The two horizontal dashed lines delineate the position-velocity slice.  \textbf{(b)} Position-velocity map for the Ori~OB\,1a region. The logarithmic contours begin at -0.6 ($\sim 2\sigma$) and then proceed in steps of 0.3.}\label{orim17}
\end{figure}

\begin{figure*}
\begin{center}
\leavevmode
\epsfxsize=8.5cm
\epsfbox{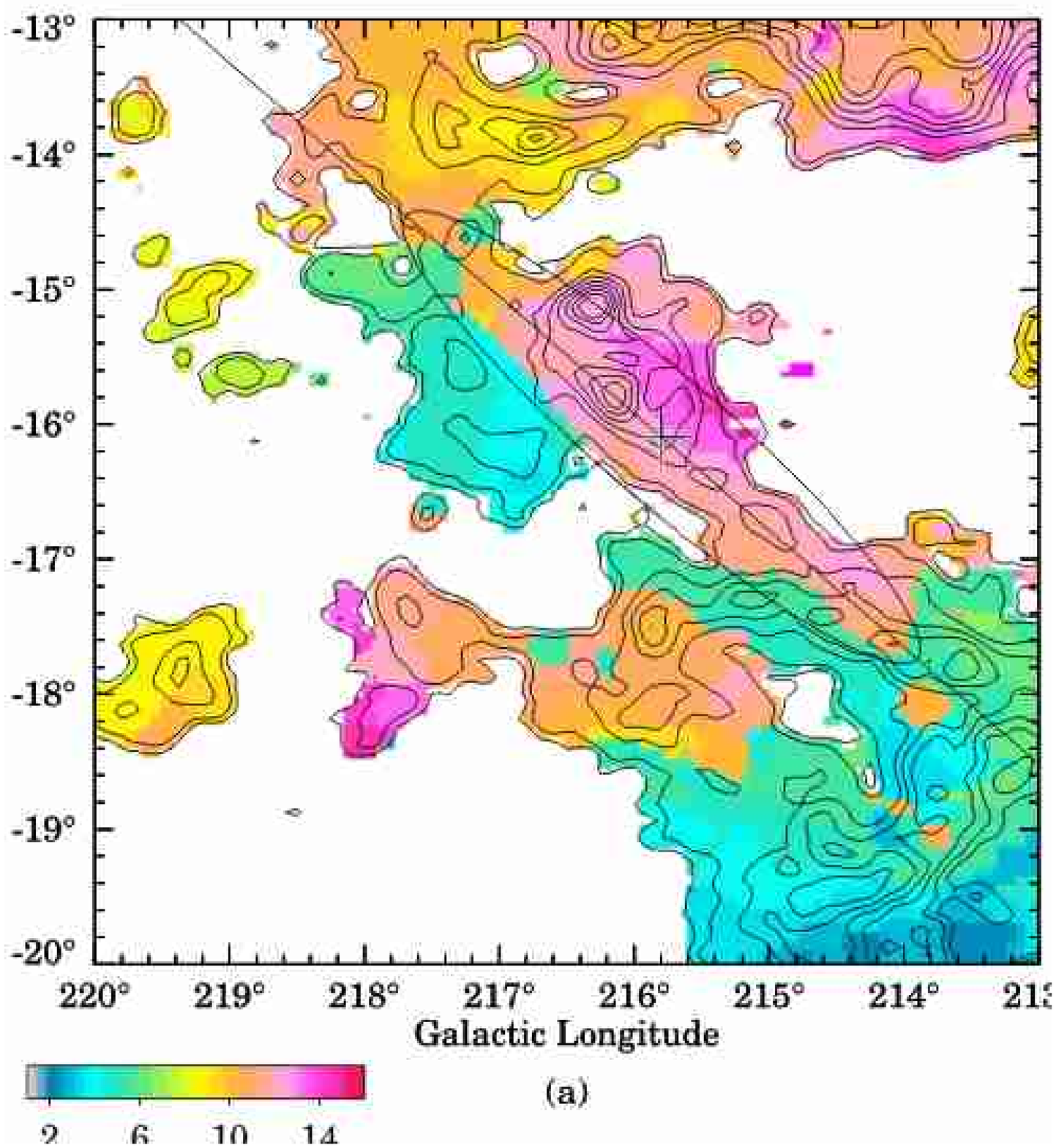}
\hfill
\epsfxsize=8.5cm
\epsfbox{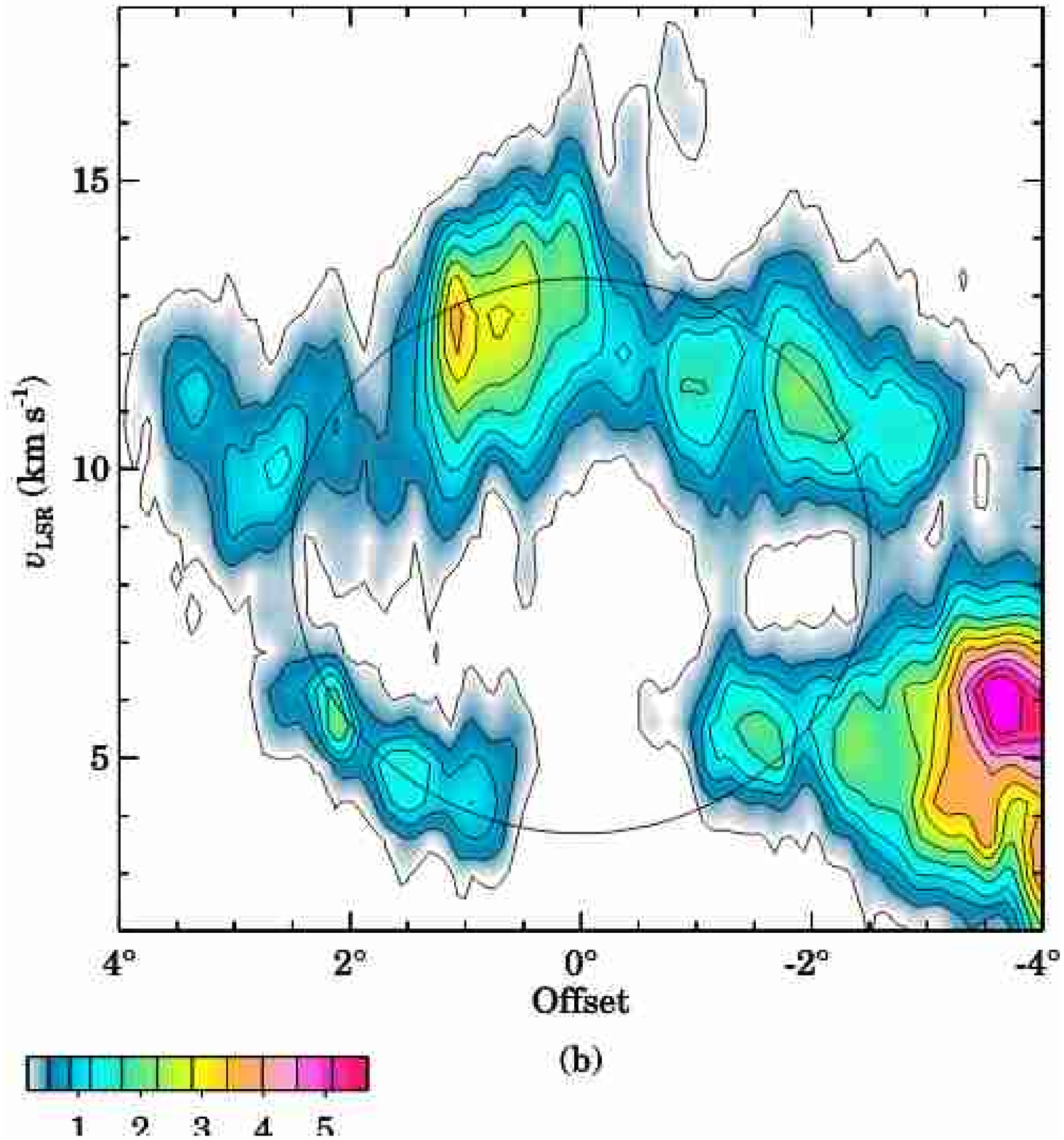}
\vspace*{2mm}
\end{center}
\caption{\textbf{(a)} Weighted mean velocity field of the NGC~2149 region. The integration range is $1 < v_{\mathrm{LSR}} < 16$\,\kms\ and the contours are at 1.7 ($3\sigma$), 2.8, 7.3, 10.1, 13.0, 15.8, 18.6 and 21.4\,K\,\kms. The solid ellipse marks the position of the expanding ring model fitted to the observations.  The solid line shows the orientation of the position-velocity slice through the ring that is presented in \textbf{(b)}, and the cross marks the location of the projected centre of the ellipse which is also the zero offset position. In \textbf{(b)} the slice was integrated over 1\fdg25, the maximum extent of the projected ellipse. The contour levels start at 0.2 with subsequent levels at 0.5, 0.9, 1.2, 1.7, 2.3, 2.8, 3.4, 4.0, 4.5, 5.1 and 5.7\,K\,arcdeg.}\label{orim18}
\end{figure*}

A possible explanation for the unusual velocity structure of the
Scissors is suggested by an examination of the kinematics of the
surrounding gas.  Fig.~\ref{orim17} presents a position-velocity map
for a wide ($\sim 6\degr$) strip which includes the Scissors, the low
longitude tip of the Orion~A cloud and a number of small cometary
clouds.  Panel (a) shows the velocity-integrated CO map
rotated 55\degr\ clockwise about the position $l =
206\degr$, $b = -17\degr$. The integration range for the
position-velocity plot is shown by the two horizontal
dashed lines. The positions of OB stars from the Ori~OB\,1a and
1b subgroups (Brown et al 1995)\nocite{Brown95} are plotted with the
same symbols as in Fig.~\ref{orim1}.  The $\mathrm{W_{CO}}$ map is
presented in this way so that a direct comparison with the
position-velocity plot, panel (b), can be made.  The kinematics of
these clouds indicate that the Scissors forms part of a semi-circular
structure that is consistent with expansion about the Orion OB\,1a
subgroup. The radius of the ring is $\sim 7\degr$, which corresponds
to a radius of $\sim 50$\,pc at a distance of $\sim 400$\,pc.  An
expansion velocity of 8\,\kms\ indicates that the ring is $\sim
6$\,Myr old, approximately half the age of the OB\,1a subgroup
(Blaauw et al. 1991).\nocite{Blaauw91}

In this model of an expanding ring the high velocity ($\sim 12$\,\kms)
clouds, characterized by the tip of Orion~A, represent the most
distant part of the ring and the low velocity gas ($\sim -1$\,\kms) in
the Scissors and the globule at $x = 8\degr$, $y = -5\degr$ in
Fig.~\ref{orim17}\,(a) represents the closest part of the ring.
This is consistent with the model used to explain the kinematics of
Orion East, but the actual distances derived using the constraining
star technique indicate that the distance between the nearest and
furthest part of the ring is $\sim 300$\,pc.  The distance estimates
may be in error, as they are based upon only a small number of stars,
but it is also possible that the ring is older than 6\,Myr and that
because of its orientation its actual size is underestimated.

\begin{figure*}
\leavevmode
\epsfxsize=8.5cm
\epsfbox{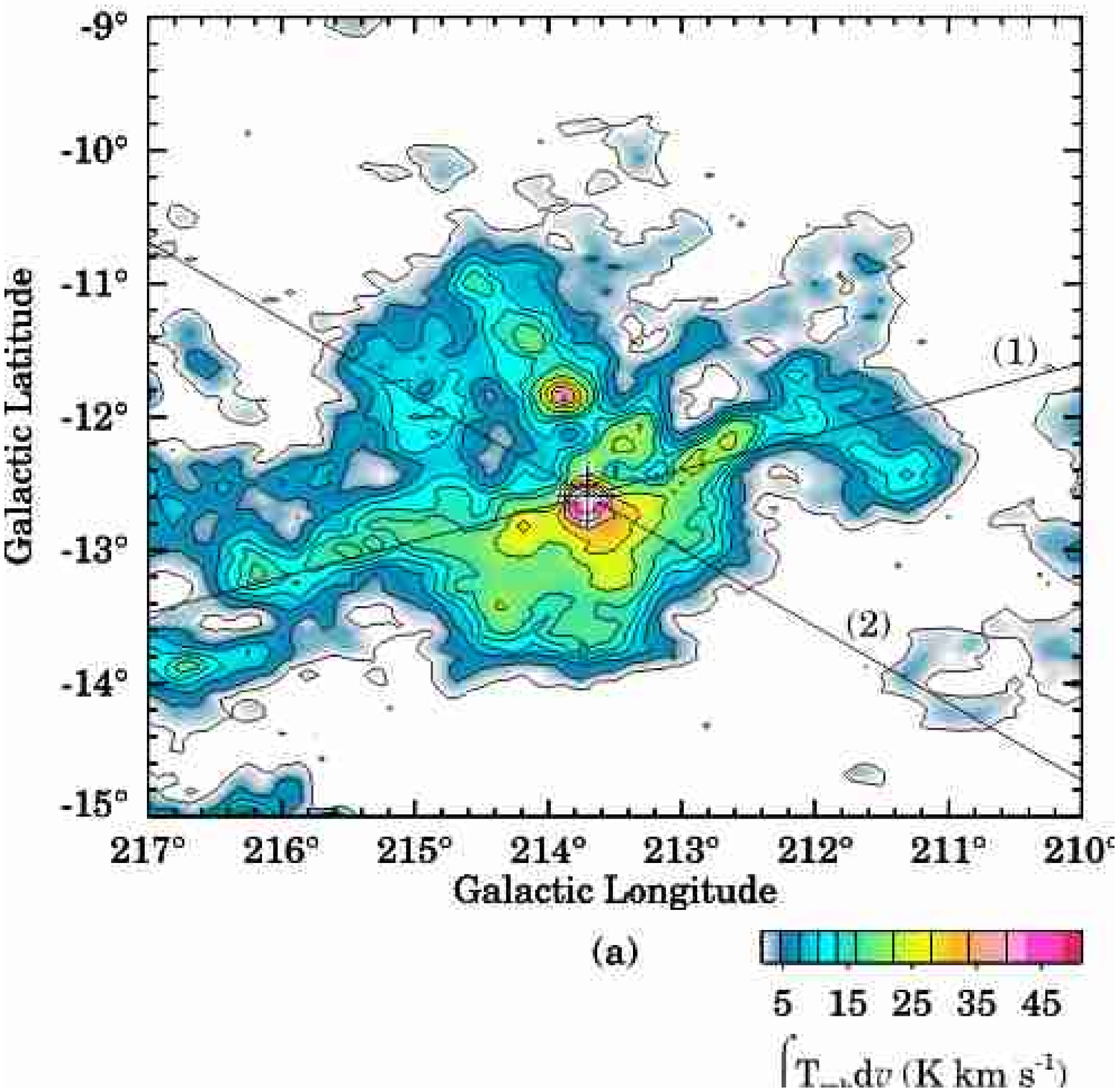}
\hfill
\epsfxsize=8.5cm
\epsfbox{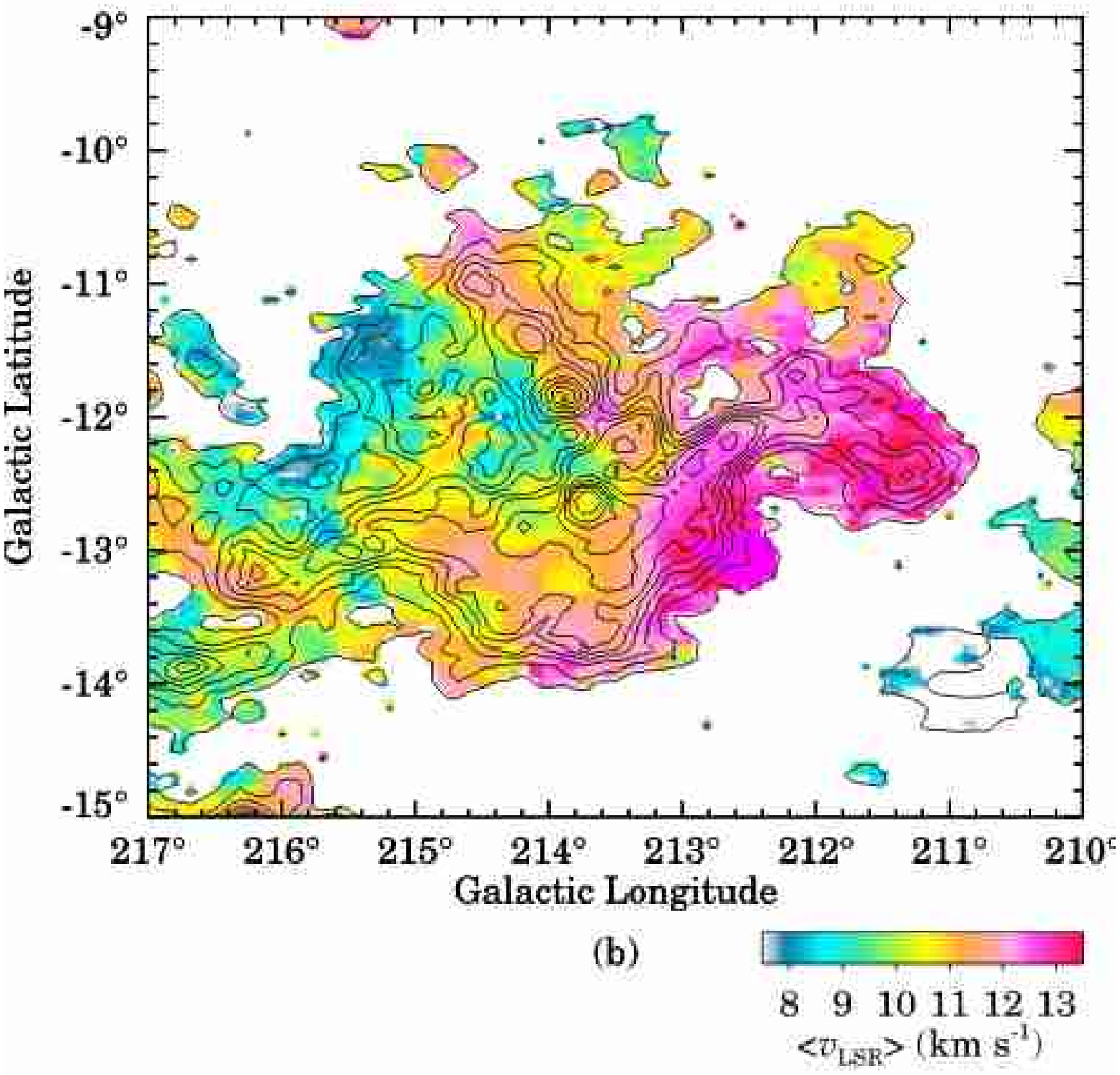}
\caption{Mon~R2 in CO. \textbf{(a)} \wco\ map of Mon~R2.  The integration range is $5 < v_{\mathrm{LSR}} < 20$\,\kms\ and the contours are at 1.7 ($3\sigma$), 4.7, 7.6, 10.5, 13.4, 16.3, 22.1, 27.9, 33.7, 39.6 and 51.2\,K\,\kms. The two solid lines mark the position velocity slices in Fig.~\ref{orim22}. The cross at $l=213\fdg71,\ b = -12\fdg59$ marks the position of the Mon~R2 outflow. \textbf{(b)} Mon~R2 emission weighted mean velocity field with \wco\ contours superimposed.}\label{orim20}
\end{figure*}

\subsection{NGC 2149: a giant expanding ring.}\label{ring}

Between the Mon R2 cloud and the high longitude end of the Orion~A
cloud are a group of molecular clouds associated with the reflection
nebula NGC 2149 (van den Bergh 1966).  \nocite{vandenBergh66}These
clouds are at a significantly higher velocity ($\sim
12$\,km\,s$^{-1}$) than those in Orion~A ($\sim 5$\,km\,s$^{-1}$).  In
most projections these clouds appear to be unconnected. However, a
three dimensional display of the data with velocity as a third
dimension indicates that the clouds in this region may be connected to
form a ring.  This structure is on a much larger scale than the
expanding shells located within Orion~A.

The structure of the CO emission from NGC~2149 was modelled as an
expanding ring.  Fig.~\ref{orim18} presents the fit of the model to
the observations. (a) is the observed peak velocity field of NGC
2149, the ellipse representing the position and orientation of the
model ring.  (b) is a position-velocity diagram along the projected
major axis of the model ring.  The two plots in Fig.~\ref{orim18}
suggest that the model is a reasonable fit to the data.

The total kinetic energy of the ring is estimated to be
$\frac{1}{2}M_{\mathrm{ring}}v^2_{\mathrm{exp}} = 3 \times
10^{48}$\,ergs, where it is assumed that all of the gas in the NGC
2149 region is expanding, i.e.  $M_{\mathrm{ring}} \equiv
M_{\mathrm{NGC 2149}} = 12.3 \times 10^3$\,\msun. At a distance of
400\,pc the radius of the ring is $\approx 40$\,pc, which corresponds
to an expansion age of $\sim 9$\,Myr.  The size and energetics of the
NGC~2149 ring are comparable to those of the $\lambda$-Orionis ring
\nocite{Lang00}(Lang et al. 2000).  However unlike the
$\lambda$-Orionis ring, the NGC~2149 ring is not associated with a
region of excess H$\alpha$ and there are no stars comparable
to $\lambda$-Ori to drive its expansion, which may therefore 
be the result of a supernova.

\subsection{Mon R2}\label{monr2}

Mon~R2 is a massive star-forming cloud between Orion~A and the
Galactic plane. Its distance is claimed to be approximately 800\,pc,
significantly greater than that of Orion~A or Orion~B, but this
estimate is quite uncertain. The constraining star technique provides
only very rough upper and lower limits for the distance to the cloud,
placing it only in the range 600--1100\,pc.  This estimate is
consistent with the photometric determinations of 830$\pm$50 (Racine
1968; Herbst \& Racine 1976) and 950\,pc (Racine \& van den Bergh
1970) \nocite{Racine68}\nocite{Herbst76}\nocite{Racine70}. Mon~R2
subtends an area of $\sim 15.3$\,deg$^2$ ($\sim 3000\,\mathrm{pc^2}$)
and has a mass of $\sim 0.75 \times 10^5$\,\msun. Both the
size and mass are comparable to those of Orion~A.  It is also located
at the same distance below the Galactic plane, $\sim 170$\,pc, and it
has a similar velocity gradient that runs roughly parallel to
the Galactic plane and in the opposite direction to that expected from
Galactic rotation (Hughes \& Baines 1985).

Mon~R2 was first mapped in CO by Kutner \& Tucker (1975),
\nocite{Kutner75}then by Maddalena et al. (1986)
and, at higher resolution, by Xie \& Goldsmith (1994)\nocite{Xie94}
with the FCRAO 14\,m telescope.  Mon~R2 has also been surveyed in
$^{13}$CO with the AT\&T Bell Laboratories 7\,m telescope (Miesch \&
Bally 1994).  \nocite{Miesch94}The high resolution surveys, and a
smaller-scale multiline study of the high-density central core
\nocite{Choi00}(Choi et al. 2000), have shown that Mon~R2 has a
unusual internal structure where most of the high density gas
has been cleared from the central core and is constrained to a
$\sim 30$\,pc hemispherical shell.  CO observations have also revealed
the presence of two outflows; a large 200\,\msun\ molecular outflow
(Loren 1981; Wolf, Lada \& Bally 1990; Meyers-Rice \& Lada
1991)\nocite{Loren81}\nocite{Wolf90}\nocite{Meyers-Rice91} that
originates from the main core and a much smaller outflow associated
with the GGD\,12--15 core \nocite{Little90}(Little et al. 1990).

Fig.~\ref{orim20}\,(a) shows the integrated Mon~R2 CO emission from the
present survey. Mon~R2 apparently contains two main structures: most
of the emission comes from a pronounced spur that extends
from the central core toward higher latitudes and the Galactic plane, or
from the low-latitude side of the cloud that forms the roughly hemispherical shell  (most clearly seen above the 16\,K\,kms\ contour) identified by Choi et al. (2000).
Extending along the spur are three discrete knots of emission;
the largest one, which is also the one closest to the main core, contains
the GGD\,12--15 core and associated outflow.

\subsubsection{The Mon~R2 shell}

\begin{figure}
\leavevmode
\epsfxsize=4.2cm
\epsfbox{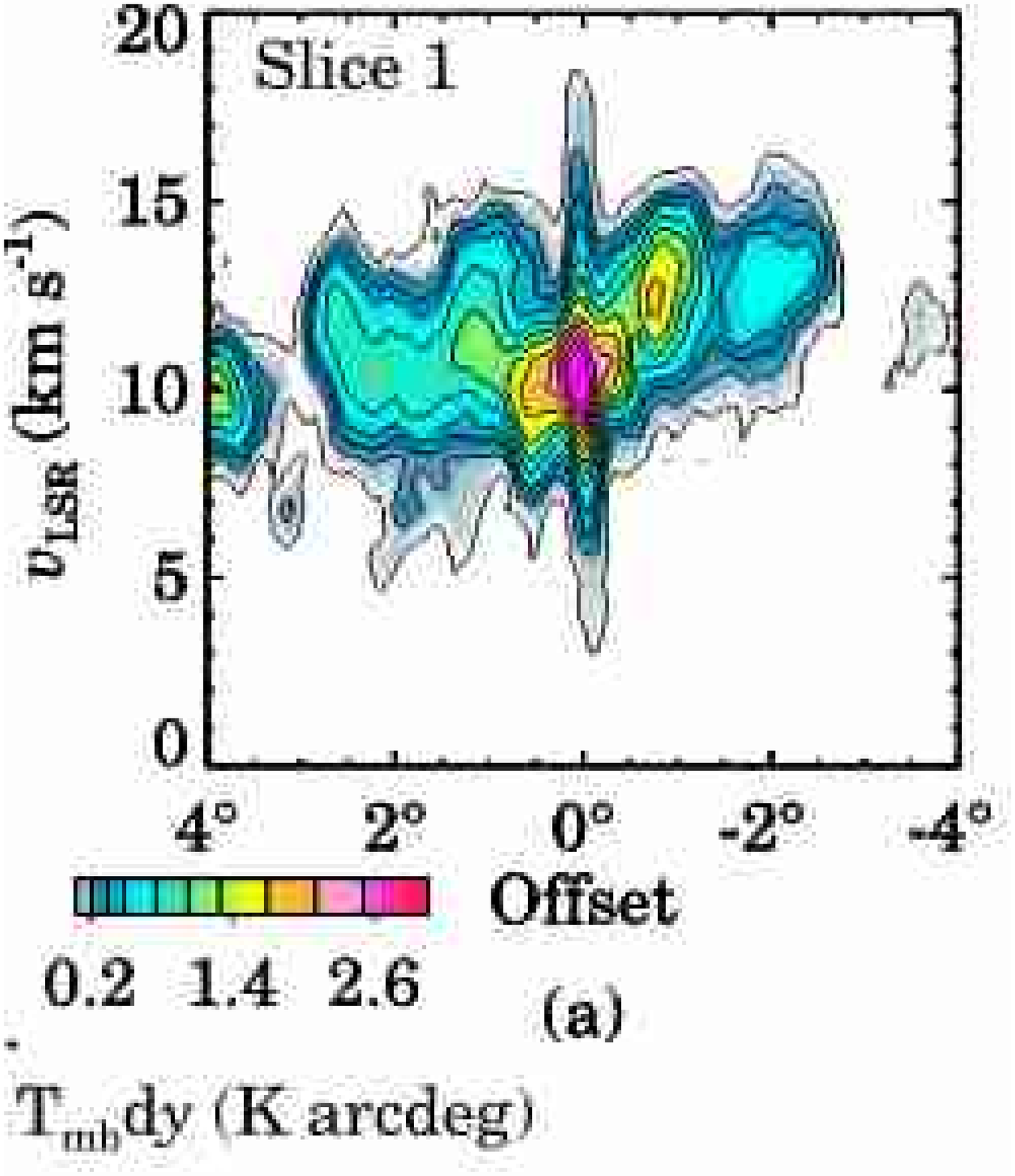}
\hfill
\epsfxsize=4.2cm
\epsfbox{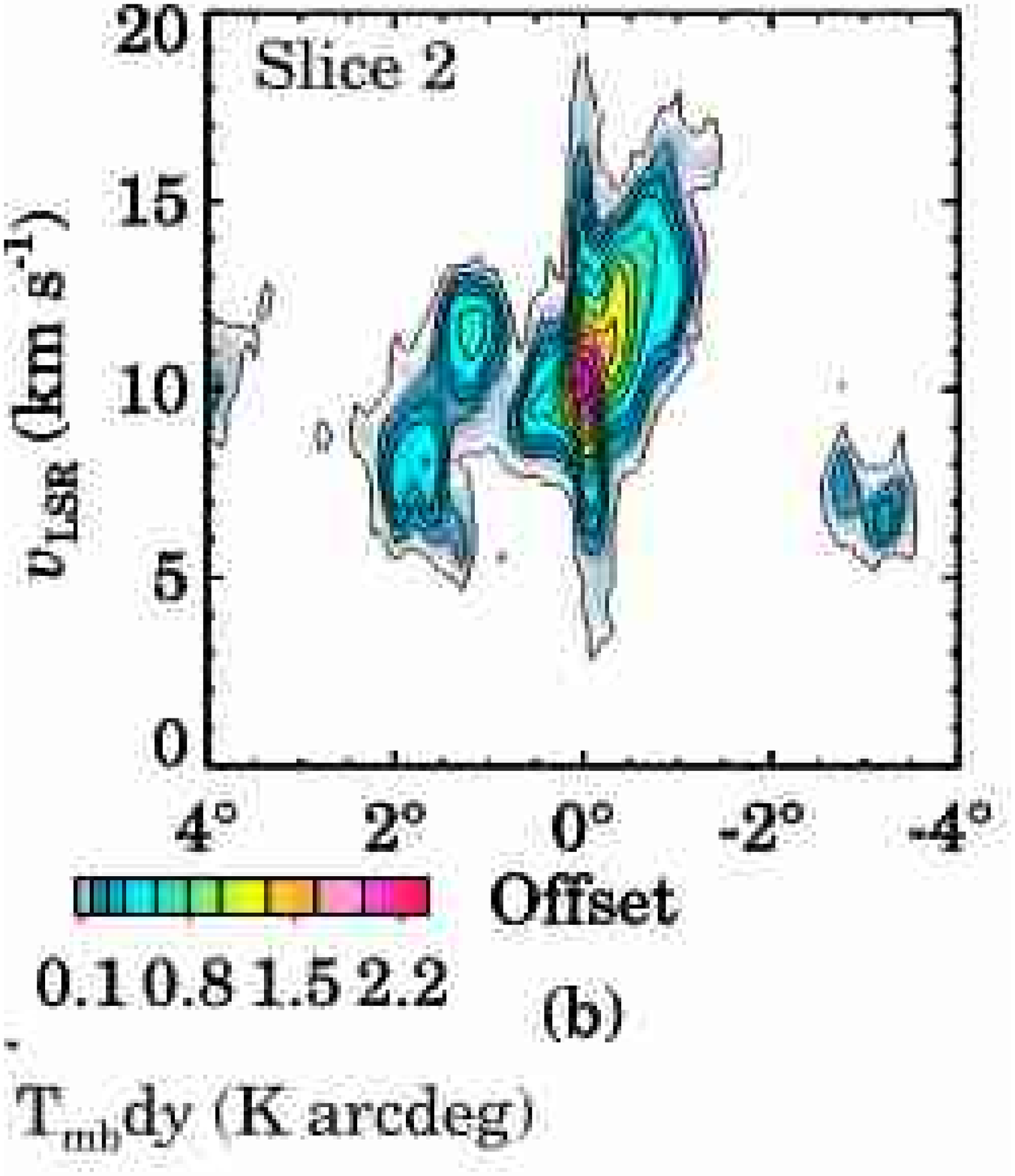}
\caption{Two position-velocity slices through Mon~R2.  The positions and orientations of the slices 1 and 2 are shown in Fig.~\ref{orim20}\,(a).  Each slice was integrated over $\sim 0\fdg25$.  Contour levels are indicated by the colour bars.}\label{orim22}
\end{figure}

Xie \& Goldsmith (1994) reported that a large fraction of the molecular
gas in Mon~R2 is confined to a large ($\sim 30$\,pc) hemispherical
shell, the projected centre of which is close to the NGC~2182
reflection nebula.
This shell is obvious in the two CO position-velocity slices in
Fig.~\ref{orim22}, because they have been integrated only over the
position of the core and not the whole cloud.  The slices go through
the position of the main outflow ($l =213\fdg7$, $b = -12\fdg6$) and
are orientated as shown by the solid lines in Fig.~\ref{orim20}(a).
In Fig.~\ref{orim22} the main lobe of the outflow appears as a
distinctive spike.  Slice (1) runs roughly perpendicular to the major
axis of the outflow, along the boundary between the low and high
emission sides of Mon~R2.  Fig.~\ref{orim22}\,(a) is dominated by the
outflow spike but the semi-circular velocity structure that
characterises the Mon~R2 shell is also apparent.  Slice (2) runs along
the major axis of the outflow, as reported by Meyers-Rice \& Lada
(1991);\nocite{Meyers-Rice91} the associated position-velocity plot is
shown in Fig.~\ref{orim22}\,(b). In addition to the outflow spike
there is a large velocity gradient along the slice which characterises
the Mon~R2 shell. The direction of the velocity gradient is not
oriented in the same direction as the outflow, which accelerates gas
at negative offsets toward lower velocities.  This suggests that the
outflow is not responsible for the shell. Furthermore, it is clear
from Fig.~\ref{orim22} that the main core is a coherent
part of the velocity structure of the shell, and is therefore likely
to be located within it.

\subsubsection{The Mon~R2 spur}

Xie \& Goldsmith (1994) \nocite{Xie94}suggested that the sharp edge of
the spur was evidence of recent formation, plausibly by shock
compression.  The corrugations that appear along the spur are also
consistent with the kinematic instability of shock front development
\nocite{Elmegreen92}(Elmegreen \& Combes 1992). Since the main Mon~R2
outflow at $l=213\fdg7,\ b = -12\fdg6$ is the most conspicuous energy
source in the cloud, it is worth asking if the spur is related to
it.

The spur originates from a position very close to the main Mon~R2
outflow (marked as a cross in Fig.~\ref{orim20} [a]), and it is
conceivable that gas entrained by the outflow has been aligned along
the spur by some mechanism.  However, for any mechanism to be credible
it must also explain two significant observations.  First, the
orientation of the outflow, as determined by Meyers-Rice \& Lada
(1991), differs by $\sim 45\degr$\ from the orientation of the spur,
and second, the base of the spur is at least 0\fdg5 from the outflow.
Jets driving smaller outflows have been observed to precess (e.g. HH
315 Arce \& Goodman 2002), \nocite{Arce02}and this mechanism could
explain the observed differences between the orientation of the clumps
and the outflow, and the offset between the base of the spur and the
outflow.  Nevertheless, the length of the spur ($\sim 30$\,pc) may be
too large for this to be realistic.  Wolf, Lada, \& Bally (1990)
\nocite{Wolf90}estimated the age of the outflow to be $1.5 \times
10^5$\,yr, so for the outflow to have swept the most distant clump out
to 30\,pc the gas would have had to have a velocity of 300\,\kms.
Although this is comparable to the speed of the jet, it is much faster
than the speed of the gas entrained by the outflow and so it is
therefore more likely that another, older, mechanism has aligned the
spur.

The projected centre of the shell
($l = 213\fdg9$, $sb = -12\fdg2$) is about halfway between
the two dense cores (the main core and the GGD\,12--15 core) that
contain the two main Mon~R2 outflows. When compared
with the centroid of the Mon~R2 shell the spur appears to be
cometary, with a well defined rounded head centered on the
GGD\,12--15 core and a tail that extends toward low latitudes.
Xie \& Goldsmith (1994) estimated the dynamical age of the shell
to be $\sim 4 \times 10^6$\,yr and its total kinetic energy to be
$5\ndash20 \times 10^{48}$\,ergs.  A similar amount
of energy deposited over about the same length of time might
produce the observed structure of the spur.

\begin{figure*}
\leavevmode
\hfill
\epsfxsize=8.5cm
\epsfbox{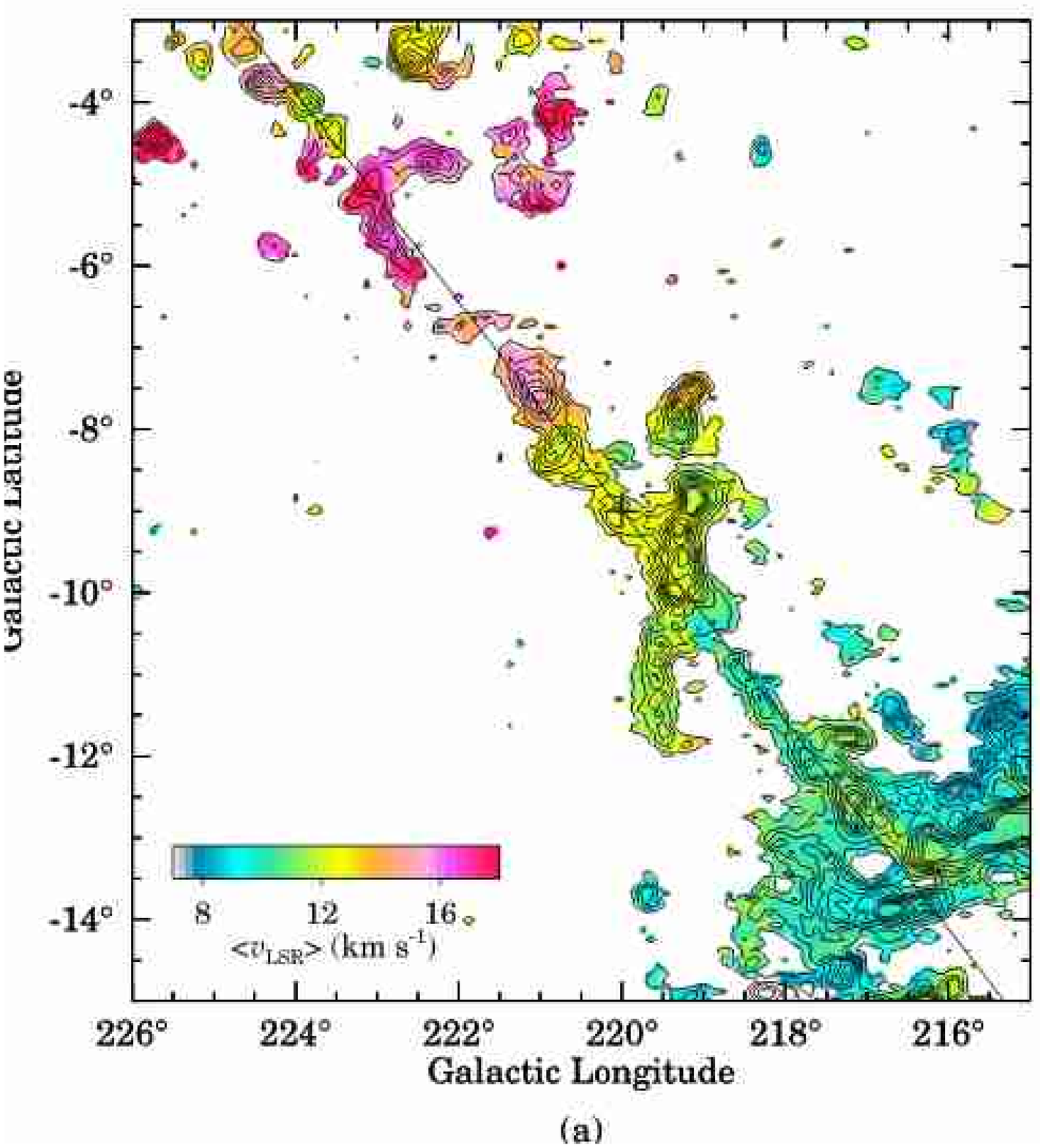}
\hfill
\epsfxsize=8.5cm
\epsfbox{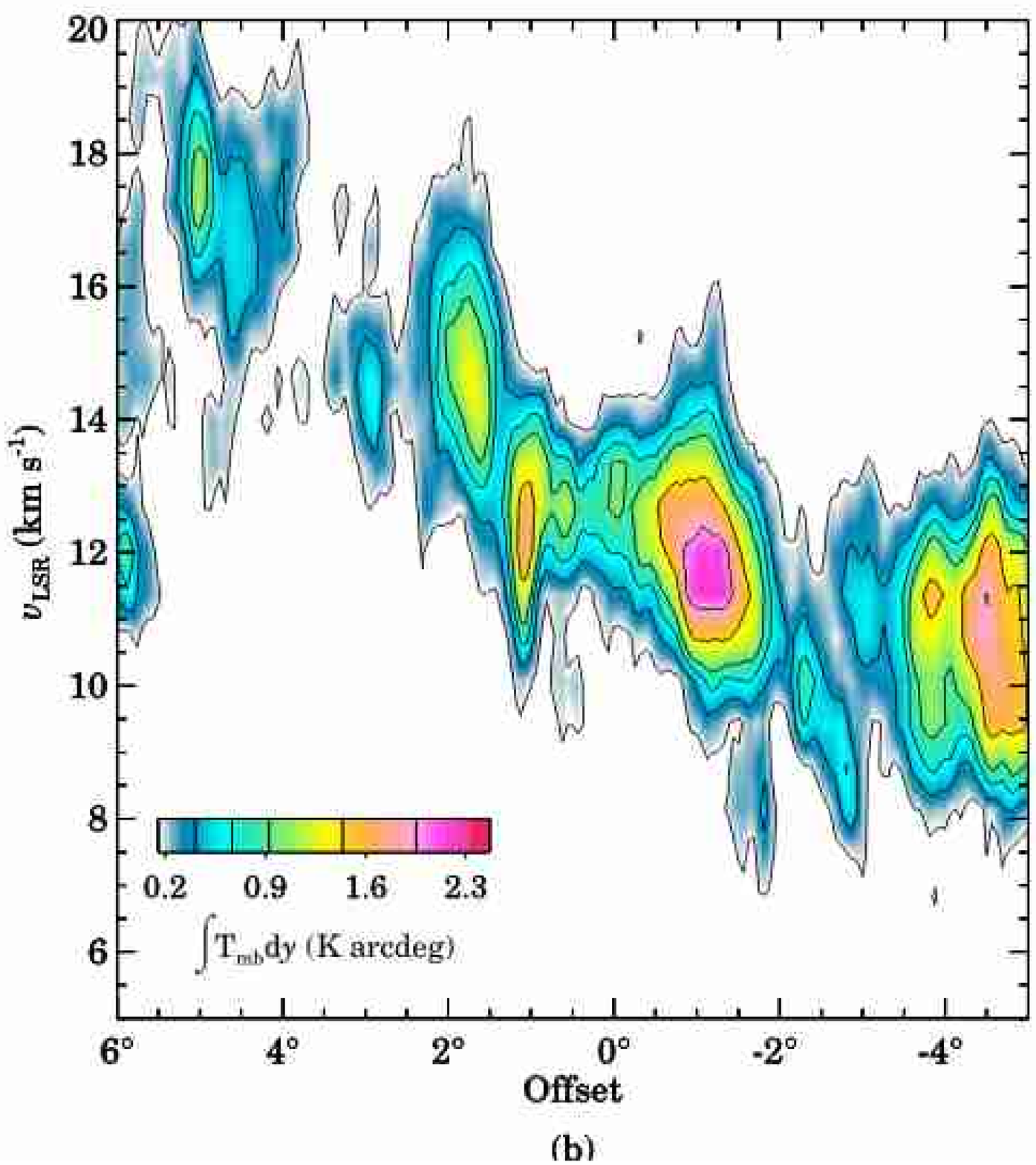}
\hfill
\caption{\textbf{(a)} Emission-weighted mean velocity map of the Southern Filament with \wco\ contours superimposed.  Contours start at 1.69\,K\,\kms\ ($3\sigma$) with subsequent steps every 1.69\,K\,\kms. The solid line shows the orientation of the position-velocity slice presented in (b), with the cross marking the position of zero offset.  \textbf{(b)} Position-velocity plot for the Southern Filament.  The data was integrated over a $\sim 1\degr$\ band centered on the filament. Contour levels are indicated by the colour bar.}\label{chpt6_56}
\end{figure*}

\subsubsection{Star-formation in Mon~R2}

Mon~R2 is a very active region of star-formation with two main
condensations centered on the positions of the largest bipolar
outflows (Shimmins, Clarke \& Ekers 1966; Beckwith et al 1976;
Thronson et al. 1980; Cohen \& Schwartz 1980; Little, Heaton \& Dent
1990).\nocite{Shimmins66}\nocite{Beckwith76}\nocite{Thronson80}\nocite{Cohen80}\nocite{Little90}
Herbst \& Racine (1976)\nocite{Herbst76} suggested that there have
been two distinct phases of star-formation in Mon~R2, the first of
which produced the many A- and B- type stars and associated reflection
nebulosity in the cloud. This phase occurred at least $6 \times 10^6$
years ago, consistent with the dynamical age of the hemispherical
shell (Xie \& Goldsmith 1994).  The second period started
10$^4$--10$^5$\,yr ago (Beckwith et al. 1976; Loren 1977; Xie \&
Goldsmith 1994)\nocite{Beckwith76}\nocite{Loren77}\nocite{Xie94} and
it is continuing within the dense cores and shock compressed ridges
that are associated with the bipolar outflows, compact HII regions,
H$_2$O masers and \emph{IRAS} point sources of the cloud.  The
correlation between the location of star-formation in Mon~R2 and the
dense gas that was shock-compressed by the previous generation implies
triggered star-formation.

\subsection{The Southern Filament and the Crossbones}

Extending from close to the Mon~R2 cloud to the Galactic plane is the
Southern Filament, a remarkably straight and narrow molecular cloud,
more than 10\degr\ in length and generally less than 1\degr\ wide.
The filament has a relatively smooth gradient characterized
by higher velocities at higher latitudes.  The velocity gradient can
be seen in the emission-weighted mean velocity map and
position-velocity slice, Fig.~\ref{chpt6_56}\,(a) and (b). The
location and orientation of the position-velocity slices is shown
by the solid line in Fig.~\ref{chpt6_56}\,(a), and its nominal centre
by the cross at $l = 220\degr$, $b = -9\degr$.

Distances were determined for two parts of the Southern Filament, the
regions above and below $b \sim -7\degr$; both found to be
at $\sim 460$\,pc.  However, these estimates were based upon only
three and five stars respectively and consequently have large relative
errors (see Table~\ref{dist}).  Despite the uncertainty in the
distance estimate it seems more likely that the Southern Filament is
related to the NGC~2149 ring and high longitude end of Orion~A than
Mon~R2.  However, there is not a clear boundary between the Southern
Filament and Mon~R2 in either projection or position-velocity space.
Possible reasons for the kinematic similarity between the gases at
different distances are explored in \S\ref{large}.

\begin{figure*}
\leavevmode
\hfill
\epsfxsize=4.75cm
\epsfbox{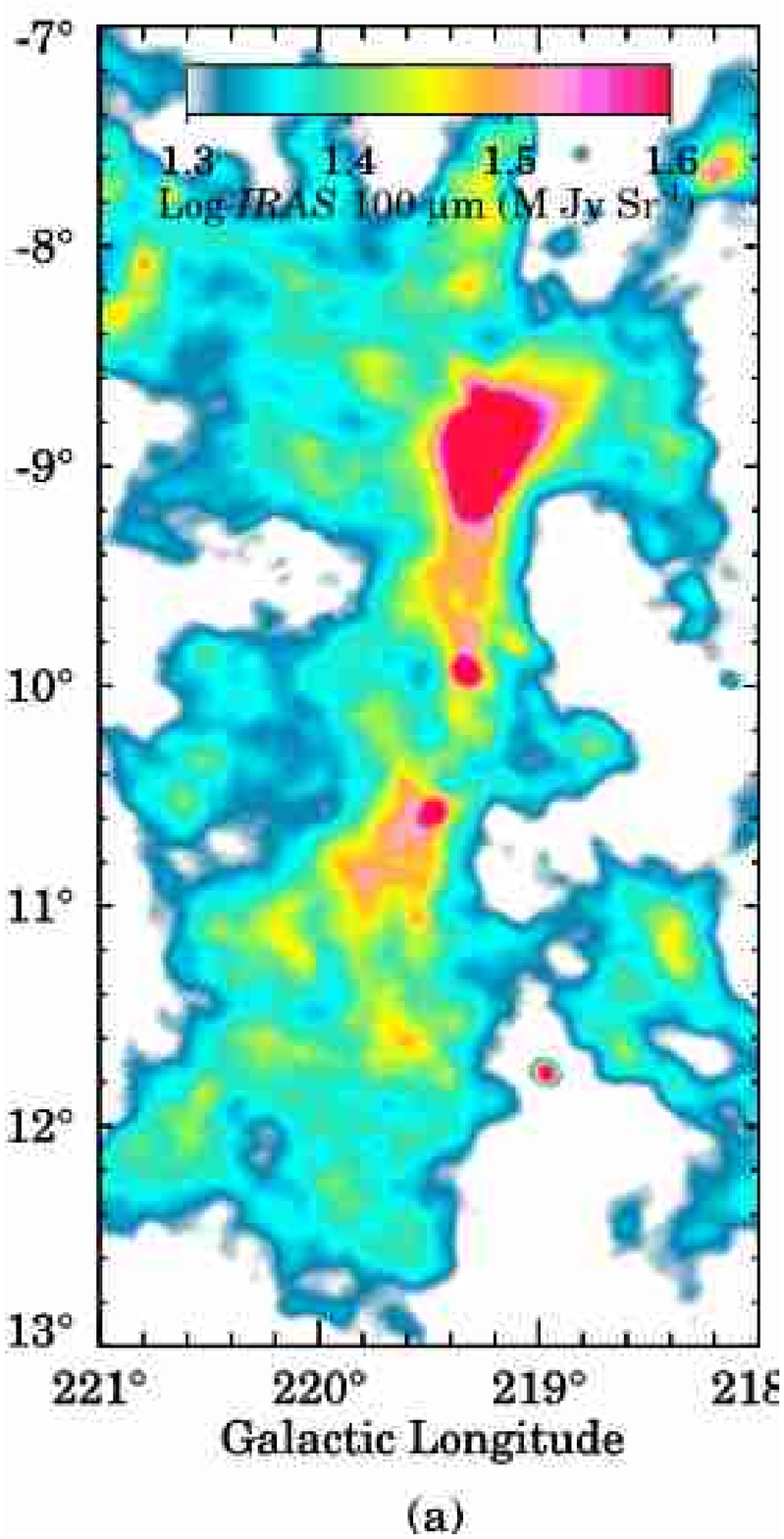}
\hfill
\epsfxsize=4.75cm
\epsfbox{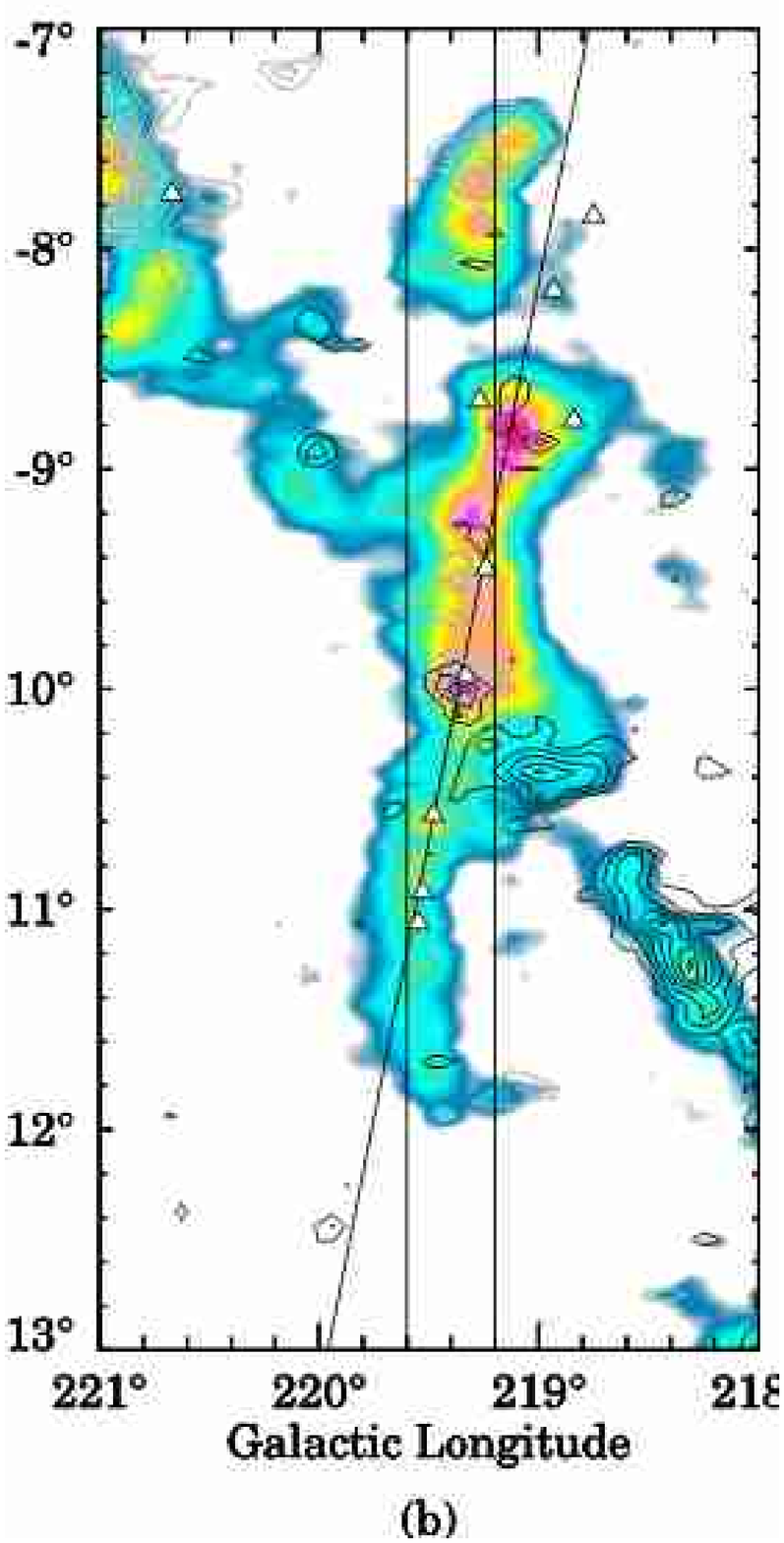}
\hfill
\epsfxsize=4.75cm
\epsfbox{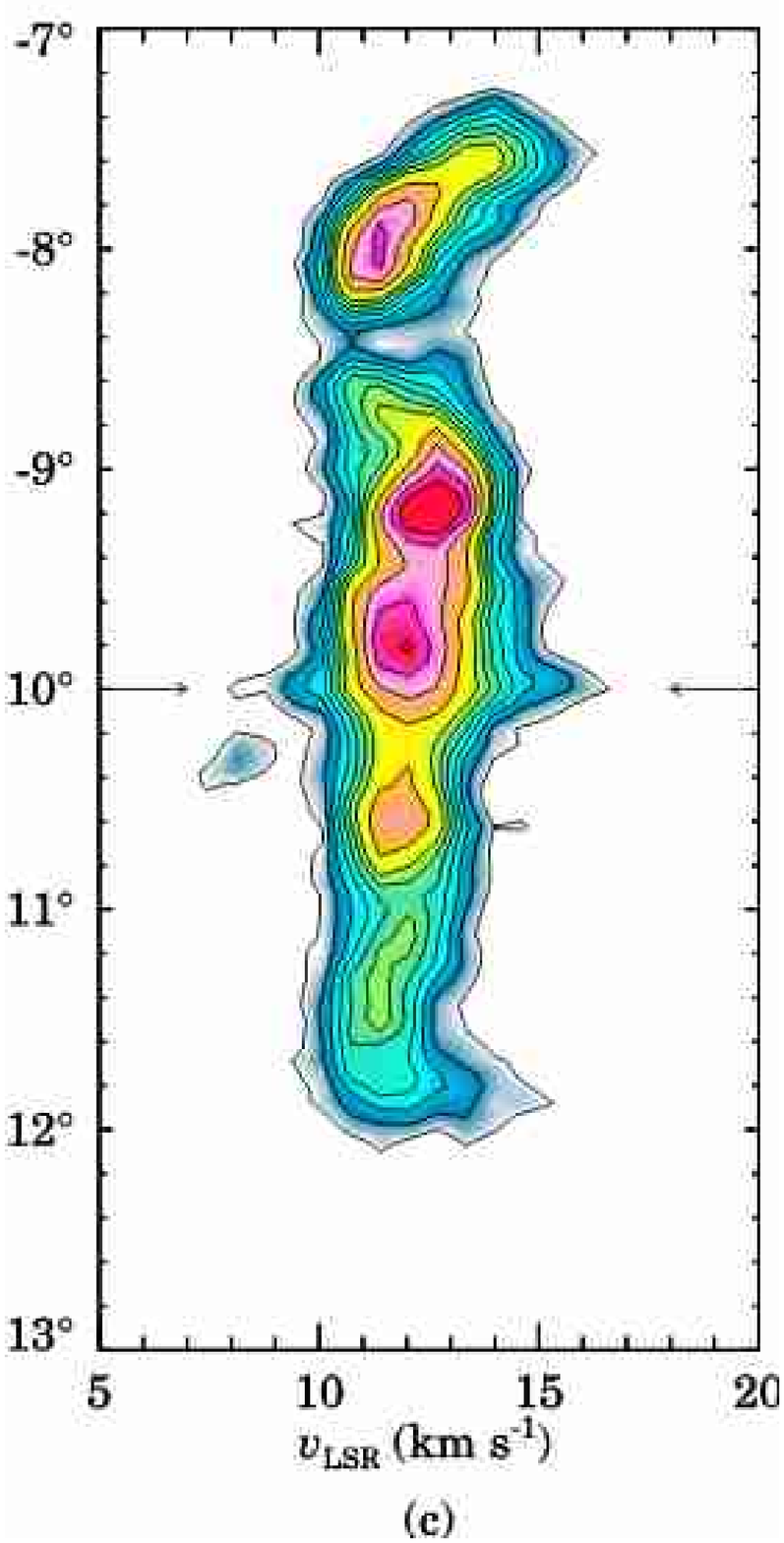}
\hfill
\caption{\textbf{(a)} Map of \emph{IRAS} 100\,$\mu$m emission from the Crossbones.  \textbf{(b)} \wco\ map of the Crossbones.  The integration range is $6.0 < v_{\mathrm{LSR}} < 18.0$\,\kms.  The white ($14.5 < v_{\mathrm{LSR}} < 18.0$\,\kms) and black ($7.0 < v_{\mathrm{LSR}} < 9.5$\,\kms) contours show the distribution of high and low velocity gas respectively. The two vertical dashed lines show the integration range for the velocity-latitude map presented in (c). The crosses mark the positions of colour selected \emph{IRAS} point sources that are potentially EYSOs. The solid line delineates the inferred major axis of the apparent bipolar outflow that is centered on $l = 219\fdg2$, $b = -10\fdg0$. \textbf{(c)} Velocity-latitude map of the Crossbones. The integration range is $219\fdg2 < l < 219\fdg6$\,\kms\ and the contours start at 0.1\,K\,arcdeg with subsequent steps every additional 0.1\,K\,arcdeg.  The arrows at $b = -10\fdg0$ mark the position of the velocity spike associated with the apparent outflow.}\label{orim58}
\end{figure*}

\begin{figure*}
\leavevmode
\hfill
\epsfxsize=8.5cm
\epsfbox{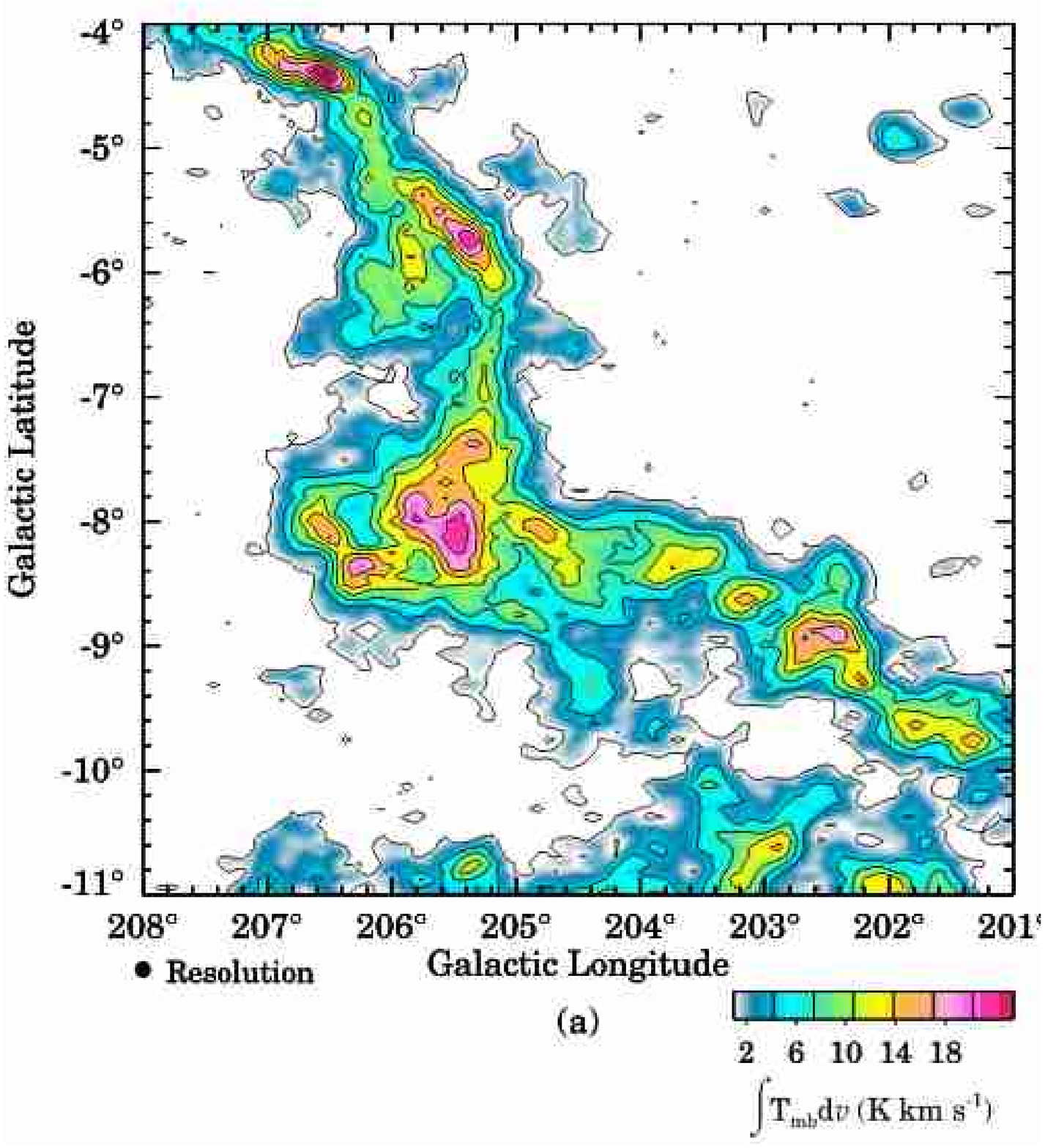}
\hfill
\epsfxsize=8.5cm
\epsfbox{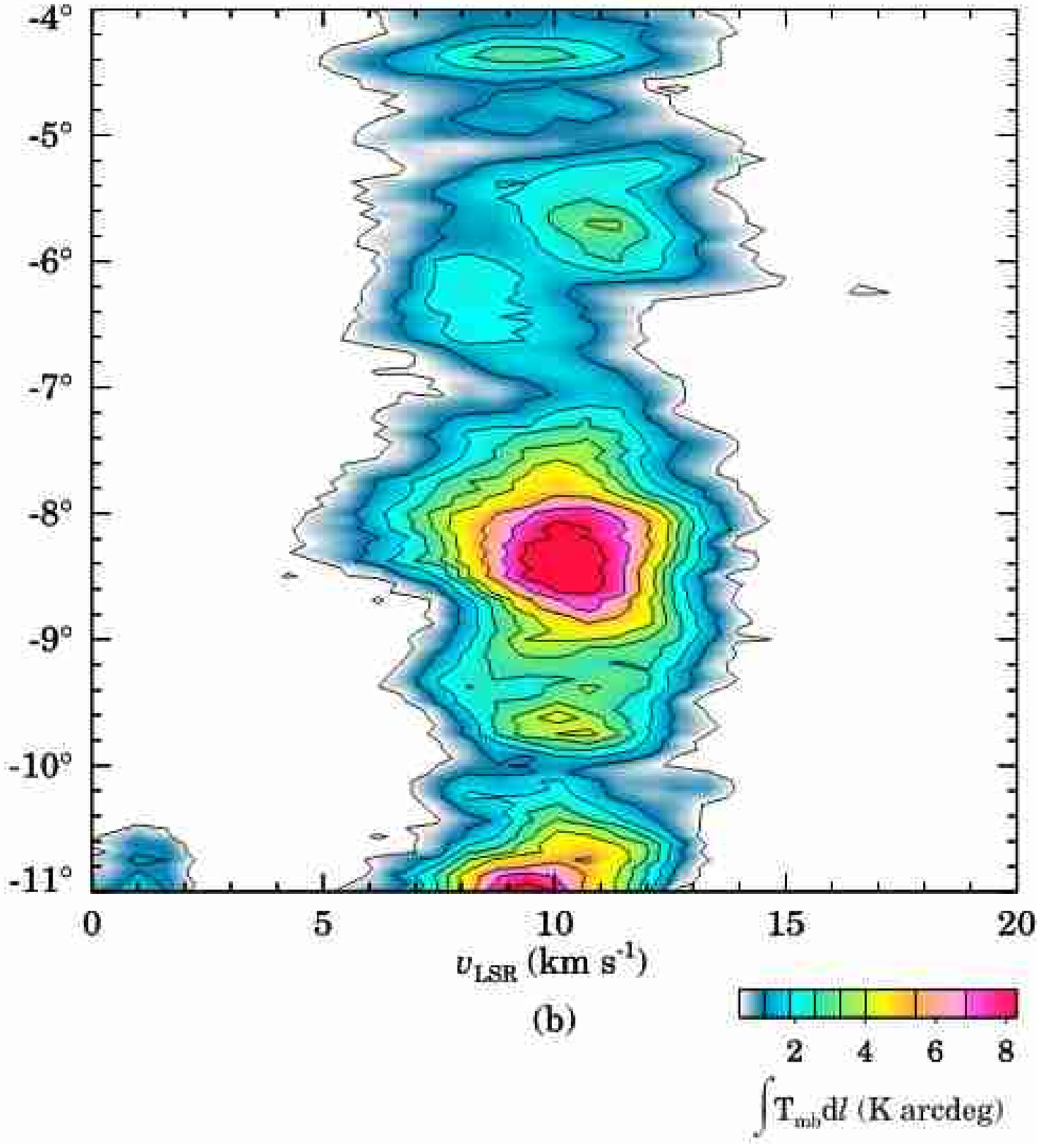}
\hfill
\caption{\textbf{(a)}\wco\ map of the Northern Filament. The integration range is $4 < v_{\mathrm{LSR}} < 16$\,\kms. The contours start at 0.97\,K\,\kms\ ($3\sigma$) with subsequent steps every additional 3.22\,K\,\kms\ ($10\sigma$). \textbf{(b)} Velocity-latitude map of the Northern Filament.  The integration range is $201\degr < l < 210\degr$ and the contours are at 0.4, 1.2, 1.9, 2.6, 3.3, 4.0, 5.4, 6.9  and 8.3\,K\,\kms.}\label{orim70}
\end{figure*}

The distinctive Crossbones clouds, first identified by Maddalena et
al. (1986), at $l \approx 219\degr$, $b \approx -10\degr$, may simply
be the superposition of two clouds along the line of sight, but the
similar kinematics of the two limbs suggests that this is unlikely.
There is some evidence, presented in Fig.~\ref{orim58}, that
star-formation is currently taking place within the Crossbones, near
the position where the two limbs intersect: (a), the
100\,$\mu$m infrared emission measured by
\emph{IRAS}, shows several knots of emission consistent with the
presence of embedded young stellar objects (YSOs).  Most promisingly
the knot at the intersection of the two limbs, $l \approx 219\fdg2$, $b
\approx -9\fdg9$ is coincident with the position of an apparent
bipolar outflow suggested by the well defined
spike in the velocity-latitude map in (c).

\subsection{The Northern Filament}

Fig.~\ref{orim70}\,(a) is the \wco\ map of a second long filament,
comparable to the Southern Filament described above, that is located
between the high-latitude end of the Orion~B cloud and the Galactic
plane.  This cloud was first mapped in CO by Maddalena et al. (1986),
who called it the Northern Filament.  However, this name is slightly
misleading because although the Northern Filament is similar in length
and orientation to the Southern Filament, it is neither as narrow nor
as straight.  It is possible that this cloud would not have been
described as a filament at all had it not been aligned parallel to the
Southern Filament.

The distance to the Northern Filament is estimated as $\sim 390$\,pc
by the constraining star technique, the same as the distance to the
high latitude part of Orion~B.  In addition to being at the same
distance, the Northern Filament and Orion~B have very similar
morphologies and kinematics along their adjacent edges, suggesting
that they were once part of a coherent structure.

The Northern Filament is relatively large and massive: it subtends an
area of 15.3\,deg$^2$ on the sky and has a mass of $0.13
\times 10^6$\,\msun. Despite its size, the cloud is
remarkably quiescent.  Molecular clouds of this size typically have
star-formation efficiencies of a fraction of one percent, which
corresponds here to $\approx 200$ young stars. There is nothing in the
Northern Filament to indicate that so much star-formation is
taking place: there are no T~Tauri stars, HII regions, reflection
nebulae, bipolar outflows or \emph{IRAS} point sources with colours
similar to those of EYSOs.

The kinematics of the Northern Filament, as summarized by the
velocity-latitude diagram, Fig.~\ref{orim70}\,(b), are fairly
unremarkable, which is not surprising considering its quiescence.
Unlike the Southern Filament, there is no large scale velocity
gradient, and the whole cloud is constrained to velocities close to
$\sim 10$\,\kms.  Part of the reason why the
velocity-latitude plot appears so smooth is because it has been
integrated over a relatively large longitude range, $200\degr < l <
210\degr$; narrower slices do reveal a more complex 
kinematic structure.  In particular there are several positions 
where the small-scale velocity structure of the cloud matches that
of Orion~B at the same longitude. 

\begin{figure}
\leavevmode
\epsfxsize=8.5cm
\epsfbox{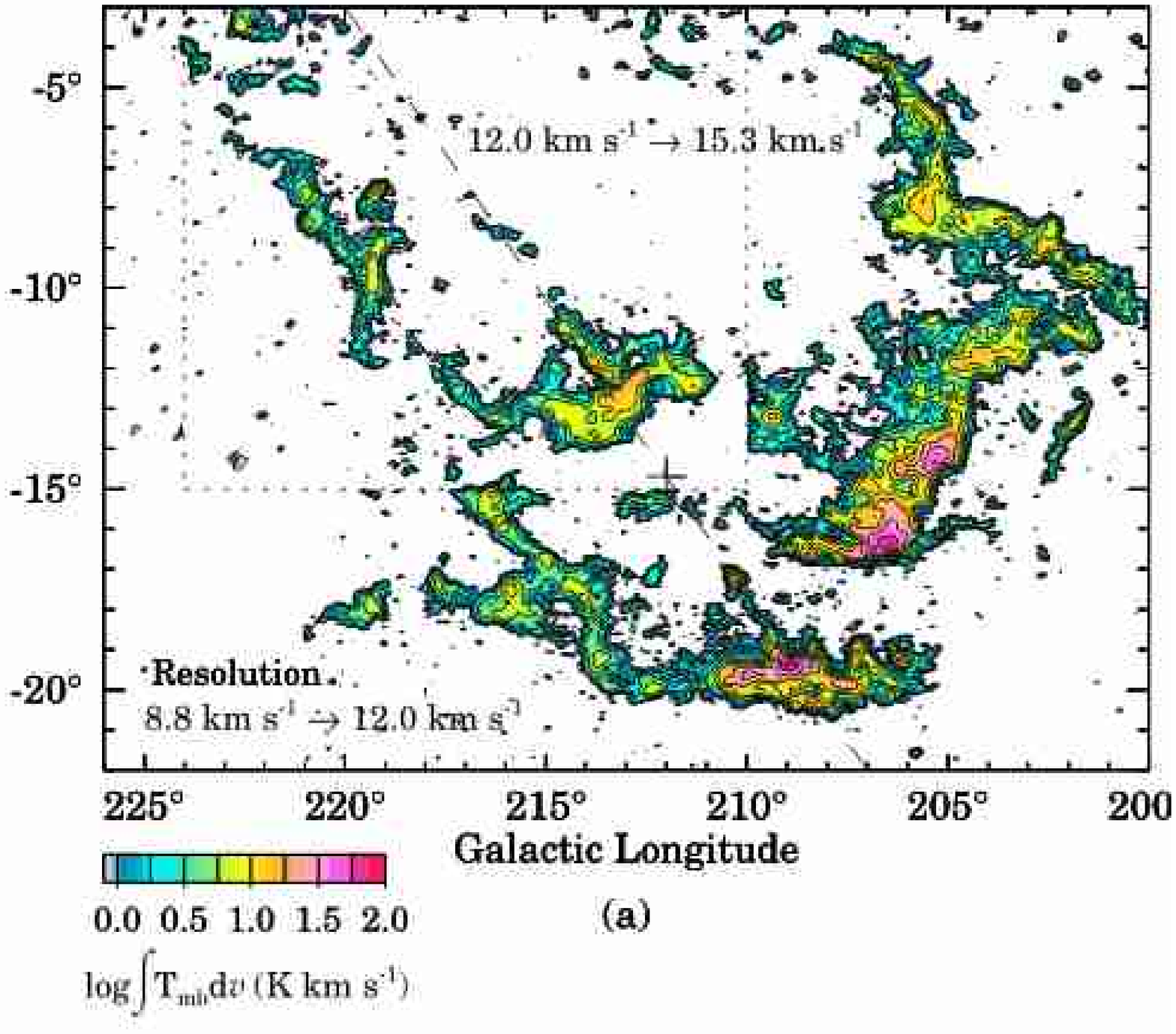}
\vspace*{0.25cm}
\vfill
\epsfxsize=8.5cm
\epsfbox{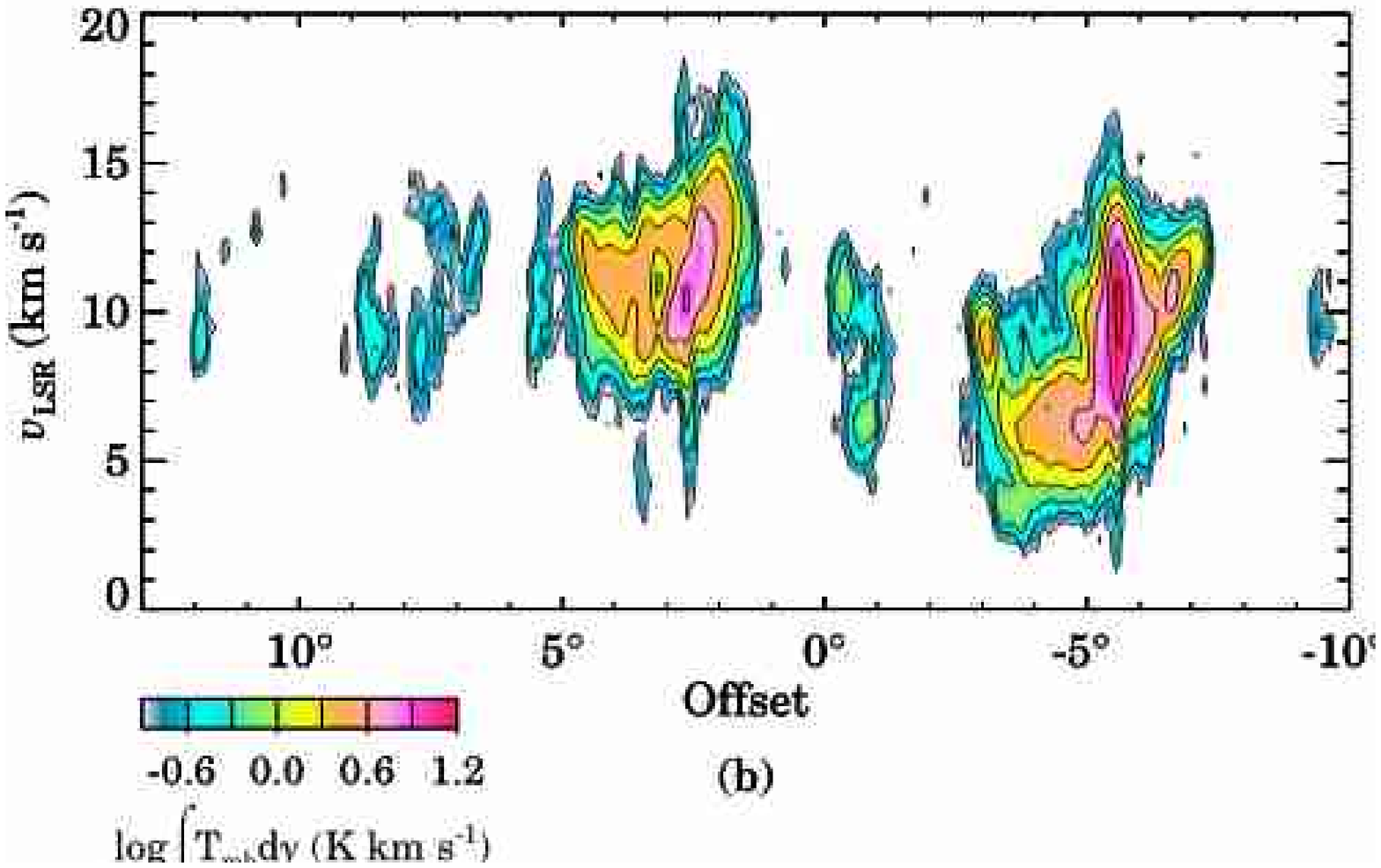}
\vspace*{0.25cm}
\vfill
\caption{\textbf{(a)} Composite \wco\ map of Orion-Monoceros. The area within the box marked by the dotted line has been integrated over the range  $12.0 < v_{\mathrm{LSR}} < 15.3$\,\kms, with the remainder of the map integrated over the range $8.8 < v_{\mathrm{LSR}} < 12.0$\,\kms. The contours are the same as the channel maps  in Fig.~\ref{orim3}. The dashed line delineates the position-velocity slice presented in (b) and the cross marks the position of zero offset. \textbf{(b)} Position-velocity slice through Orion~A and Mon~R2.  The contours start at -0.9 ($\sim 3\sigma$) and go up in steps of 0.3 (log K\,arcdeg). }\label{orim42}
\end{figure}

\section{Large-scale structure of Orion-Monoceros.}\label{large}

The similarity between the Orion~A and Mon~R2 clouds has already been
noted (Section~\ref{monr2}).  They are at approximately the same distance
from the plane, they have about the same size and mass, and they both
have large-scale velocity gradients in a direction opposite to that of
Galactic rotation.  Fig.~\ref{orim42}\,(b) is a position-velocity
slice through the emission peaks of the Orion~A and Mon~R2 clouds. The
orientation of the slice is delineated by the dashed line in
Fig.~\ref{orim42}\,(a) which shows a composite of two of the
Orion-Monoceros channel maps (Fig.~\ref{orim3}). The map is
presented in this way to emphasise that the slice is parallel to the
Northern and Southern Filaments and that it passes along emission
spurs in both Orion~A and Mon~R2 that are also parallel to the
filaments.  The morphological and kinematic
similarity between Orion~A and Mon~R2 is striking. Both clouds are
home to giant molecular outflows, as characterized by the emission
spikes at offsets of -5\fdg5 and 3\degr, both have lobes of
high-velocity gas on their low Galactic latitude sides and both have
spurs of emission that extend from their central core toward the
Galactic plane and are parallel to the Southern Filament.

Mon~R2 is more that $\sim 300$\,pc further away than Orion~A and
has not usually been considered to be part of the Orion
complex.  However, when confronted by the numerous similarities
between Orion~A and Mon~R2 it is difficult to avoid speculating that
these clouds, and by extension all the clouds in the region,
share a common origin.


Heiles (1998) proposed that the formation of Orion-Monoceros was
associated with the expansion of the Vela supershell (GSH238+00+09),
which may have originated from the cluster Collinder 121 (Cr 121)
located near $l \sim 235\degr$, $b \sim -10\degr$, $v_{\mathrm{LSR}} =
17 \pm 4$\,\kms\ \nocite{Blaauw91}(Blaauw 1991), and at a distance of
$546 \pm 30$\,pc \nocite{Hoogerwerf97}(Hoogerwerf et al. 1997). The
Vela supershell encloses an enormous irregular bubble in the region
where it has blown out of the Galactic plane, one side of which runs
approximately parallel to the Southern Filament, before becoming more
spread out at lower latitudes.

Heiles (1998)\nocite{Heiles98} claimed that the
Vela supershell follows the expansion law $R \propto t^{0.31}$,
meaning that the walls of the supershell had travelled $\sim 80\%$ of
the way to their present location at half the age.  Thus the Vela
supershell would have overrun the position of the Orion~OB association
at about the time that the OB\,1a subgroup was formed.

Bally et al. (1998) proposed that a collision of the Vela
supershell with a fossil remnant of the Lindblad Ring may have played
a role in triggering the formation of the proto Orion-Monoceros
complex. They suggest that the ISM was first compressed when it was
swept up by the Lindblad Ring supershell; as the expansion of this
superbubble slowed the dense gas in the shell was compressed
further by the impact of the Vela supershell, enhancing the tendency 
of the gas to collapse via gravitational instability and
form stars.

The structure of Orion-Monoceros close to the Orion~OB association is
now dominated by the local effects of stellar winds, supernova
explosions and the expansion of the Orion-Eridanus bubble, but
much of the underlying large-scale structure is consistent with the
formation sequence described above.  In particular the alignment of
the Southern Filament and the Orion~A and Mon~R2 spurs with the wall
of the Vela superbubble does not appear to be purely coincidental.

\section{Summary and Conclusions}\label{conc}

This paper presents the results of a new CO survey of the
Orion-Monoceros complex carried out with the Harvard-Smithsonian
1.2\,m telescope. The new observations are more sensitive and have
better sampling than the only previous large survey of the region
(Maddalena et al. 1986). Distances to the majority of the clouds in
Orion-Monoceros are determined using the constraining star technique
and the \emph{Hipparcos} catalogue.  The new distances aided the
three dimensional analysis of the molecular gas and it relationship
to the other phases of the ISM.  The main conclusions of this analysis
are as follows:
\begin{enumerate}
\item Stellar winds and ionizing radiation from the massive 
stars of the Ori~OB\,1b and OB\,1d subgroups are responsible for changes
to the morphology and kinematics of the molecular gas in Orion~A and
Orion~B today.  
\item The kinematics and three dimensional structure of 
Orion~A and Orion~B are consistent with stellar wind
driven compression centered on Ori~OB\,1b.  Knots of compressed gas
are coincident with the most active star-forming regions in Orion~A
and Orion~B, suggesting that triggering may have occurred.
\item The Orion-Eridanus superbubble of hot H$\alpha$ emitting gas
inflated by the Orion~OB association and delineated in projection 
by Barnard's loop has overrun many small clumps of dense gas.
These globules have cometary shapes caused by their interaction
with the expanding bubble, and many display signs of enhanced 
(triggered) star-formation in their heads. However, only one globule, 
Orion-East, displays a fairly unambiguous connection to the ionizing
radiation of Barnard's Loop, and this globule is significantly closer
than the main molecular clouds of the complex.  Therefore, we conclude
that Barnard's Loop delineates the closest part of the Orion-Eridanus
bubble and not the part adjacent to Orion-Monoceros.
\item  The most distant cloud in the complex, 
Mon~R2, is quite similar to the large molecular cloud associated with 
the Orion nebula that is significantly closer.
Both clouds have spurs of emission that are aligned with the
Southern Filament (and the Vela supershell).  These features suggest
that the clouds of the region share a common origin.  
\item The new CO survey supports the proposals by Bally et al. (1998)
and Heiles (1998) that the formation of the Orion-Monoceros complex
was triggered by the passage of the Vela supershell and possibly 
its interaction with a preexisting supershell associated with the
Lindblad Ring.

\end{enumerate}


\begin{acknowledgements}
Part of this work was supported by the \emph{Particle Physics and
Astronomy Research Council (PPARC)} grant number PPA/S/S/1998/02675.
The authors are grateful to E.S. Palmer for skillfully maintaining 
the 1.2\,m telescope, and to Jodie Dalton and Adam 
Kampff for their help with the observations.
\end{acknowledgements}

\bibliographystyle{aa}
\bibliography{refs}

\end{document}